\documentclass[12pt]{article}
\usepackage[utf8]{inputenc}

\usepackage{hyperref}
\usepackage{graphicx}
\usepackage{fullpage, amsmath, amsthm, amssymb, wasysym, array, bbm, latexsym}
\usepackage{natbib}
\usepackage{epsfig}
\usepackage{wrapfig}
\usepackage{mathtools}
\usepackage{rotating}
\usepackage{epstopdf}
\usepackage{caption}
\usepackage{lipsum}
\usepackage{afterpage}
\usepackage{bm}
\usepackage{hyperref}
\usepackage{subcaption}
\usepackage{pdftricks}
\usepackage{adjustbox}
\usepackage[official]{eurosym}
\usepackage{sidecap} 
\sidecaptionvpos{figure}{c} 

\usepackage{ulem} 
\usepackage{multirow}
\usepackage{hhline}

\usepackage{color}

\newlength\savedwidth

\newcommand\thickhline{\noalign{\global\savedwidth\arrayrulewidth\global\arrayrulewidth 2pt}%
\hline
\noalign{\global\arrayrulewidth\savedwidth}}

\begin{document}

\begin{center}
{\Large
\textbf{Estimating the correlation between operational risk loss categories over different time horizons}
}


\vspace{0.5cm}

{\large

Maurice L. Brown ; Cheng Ly$^{\dag}$
}
%
%
\end{center}

\begin{abstract}
    Operational risk is challenging to quantify because of the broad range of categories (fraud, technological issues, natural disasters) and the heavy-tailed nature of realized losses. Operational risk modeling requires quantifying how these broad loss categories are related. We focus on the issue of loss frequencies having different time scales (e.g., daily, yearly, monthly basis), specifically on estimating the statistics of losses on arbitrary time horizons. We present a frequency model where mathematical techniques can be feasibly applied to analytically calculate the mean, variance, and co-variances that are accurate compared to more time-consuming Monte Carlo simulations. We show that the analytic calculations of cumulative loss statistics in an arbitrary time window are feasible here and would otherwise be intractable due to temporal correlations. Our work has potential value because these statistics are crucial for approximating correlations of losses via copulas. We systematically vary all model parameters to demonstrate the accuracy of our methods for calculating all first and second order statistics of aggregate loss distributions. Finally, using combined data from a consortium of institutions, we show that different time horizons can lead to a large range of loss statistics that can significantly affect calculations of capital requirements.
\end{abstract}

\section{Introduction}

It is imperative to properly quantify risk \citep{jivrina11} since it drives financial behavior.  A particularly 
challenging risk type to model is Operational Risk, 
defined as the risk of loss resulting from inadequate or failed internal processes, people or systems, or from external events \citep{bis_2006}. 
Traditionally, operational risk has been 
neglected compared to credit and market risk; it is often thought of as ``other" despite how it has severely harmed institutions when 
not properly accounted for \citep{bis_2006,chern08book}.  
For example, in the last few decades operational losses have been extraordinary, leading to wholesale changes for 
many institutions in the form of bankruptcies, mergers, re-organizations, and other lasting damage \citep{chern08book,deFontn03}. 
The major sources of operational risk include IT security, over reliance on suppliers, 
integration of acquisitions, fraud, error, customer service quality, 
alleged wrongdoing in business practices \citep{afonso19}, 
regulatory compliance, recruitment, training and retention of staff, and social and environmental impacts \citep{bis_2006}.  
Operational risk loss data are used in models to calculate statistics that aid in decision making for the size of capital 
reserves but also to make predictions to mitigate and understand the dynamics of operational losses.  

Operational risk is important because it has accounted for a large proportion of losses \citep{afonso19}, especially for the largest 
institutions. Quantifying it presents unique mathematical and statistical challenges stemming from the broad range of 
risk categories, heavy-tailed data, as well as lack of data for certain categories 
(e.g., Disasters and Public Safety, Technology and Infrastructure Failure, see Table \ref{tbl:riskAvgs}).  
Moreover, losses in different categories occur with different frequency time-scales, ranging from a daily, monthly, or yearly basis. There are many global 
factors (business cycle, pandemics, wars, etc.) that impact multiple loss categories, leading to correlations between different categories of losses.  

There are risk categories where the data is scarce and commonly used methods to fit a loss distribution model to data would suffer from insufficient sample size. Estimating the correlation between loss categories with scarce data is challenging; the status quo is to 
calculate a time-averaged correlation/covariance of the loss data in the same time window used in a copula 
model \citep{mcneil15}. 
With scarce data, it is common to use the raw data in shorter time windows (i.e., daily or monthly) to get more 
samples rather than on cumulative data summed over the desired time window, which should be yearly data 
since the desire is to calculate yearly capital \citep{chern08book}. Thus, we focus on how different 
time horizons of losses potentially impact loss statistics.

Shedding light on how loss statistics (for a specific category and between different categories) depend on time horizons, or \textbf{time windows}, 
naturally leads to focussing on the frequency distribution.  
We introduce a stochastic point process model for the frequency distribution that has two main parameters: average frequency and a time scale parameter. The advantages of this model are as follows: i) the parameters could be fit with sufficient data but are also transparent enough to understand when fitting to scarce data, ii) when coupled with an independent severity distribution model, the auto- and cross-correlation functions are mathematically tractable, so we can analytically calculate the cumulative loss statistics over varying time windows using tools from signal processing \citep{kay1993} and 
computational neuroscience \citep{lindner05,shea2008correlation,BarreiroLy_RecrCorr_17,BarreiroLy_jmns_18}.  
Varying all model parameters, we show the advantages and shortcomings of our calculations compared to 
large-scale Monte Carlo simulations. 
Finally, using combined data from a consortium of 
institutions, we fit our model to loss statistics of an average institution.  
We show that different time windows of (cumulative) losses 
can lead to a large range of loss statistics, which in turn can significantly effect calculations of capital requirements.  

\section{Results}

A central Operational Risk Modeling objective is to estimate the cumulative loss distribution because the 99.9 percentile of this is a common capital requirement \citep{jivrina11} . 
Operational losses are realizations of a continuous stochastic process from both a frequency distribution (event occurrences) and severity distribution (amount of loss) \citep{jivrina11}. 
The Basel II Committee has suggested several approaches to calculate the capital requirements to withstand exposure to operational risk, 
with varying levels of sophistication: (1) Basic Indicator Approach, (2) Standardized Approach and (3) Advanced Measurement Approaches (\textbf{AMA}) \citep{bis_2006,chern08book}.  
Among the 
3 approaches, the AMA is the required approach for large institutions \citep{bis_2006} and known to be the most risk sensitive compared to the other 2 approaches that are 
often used for smaller institutions \citep{bis_2006,chern08book}. 
Here we will only focus on the AMA because it is used for larger banks that tend to suffer larger operational losses, and requires the most complex mathematical and statistical methods. 
Our proposed methods could potentially address the shortcomings of the AMA we discussed in the Introduction. 

The broad operational loss categories naturally have different time scales for frequency of events as well as different loss amounts (severity). 
The AMA approach entails modeling both the frequency and severity distributions, and ultimately combined into a single \textbf{aggregate loss 
distribution} for a given operational loss category \citep{chern08book}.  
The severity distributions, are often modeled by parametric distributions 
of positive random variables that are heavy-tailed; e.g., lognormal, Weibull, Pareto and Burr \citep{chern08book}.  
The frequency distribution is commonly modeled with a Poisson Process or the negative binomial distribution \citep{chern08book}. 

\begin{figure}[!htb]
\centering
\includegraphics[width=\textwidth]{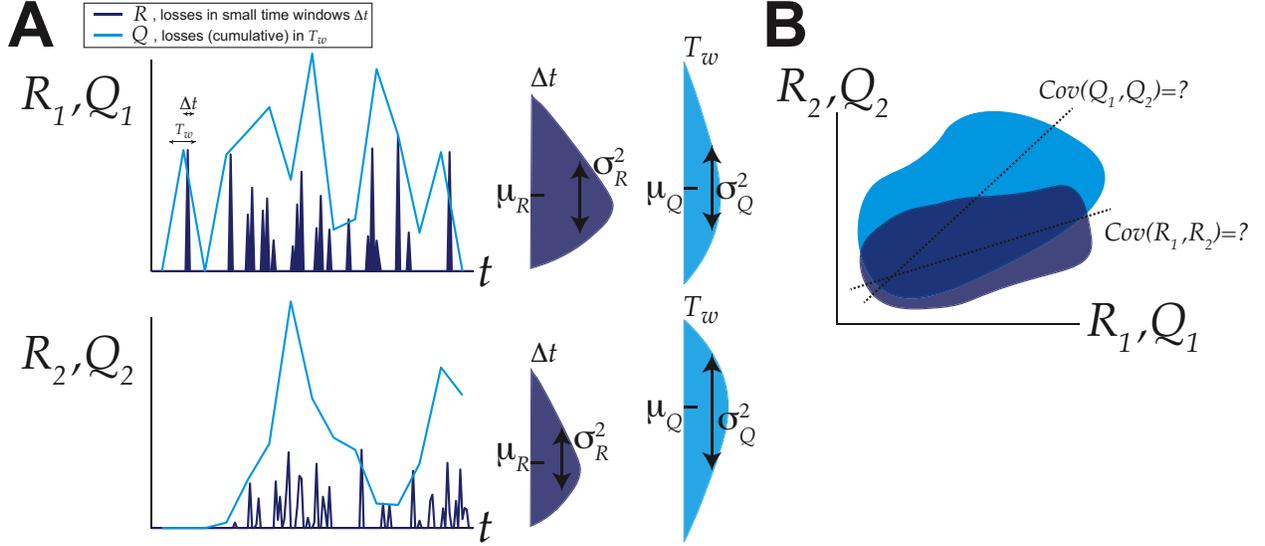}
\caption{\label{fig1:setup} \textbf{Operational risk loss statistics can vary depending on risk category 
and on time window of observations.}  \textbf{A)} Examples of two 
different loss time series (top and bottom).  For each loss category, 
two time series from the same loss data are shown: losses in small windows ($\Delta t$ dark blue) which are 
denoted by $R_j(t)$ and cumulative losses in time window $T_w$ denoted by $Q_j(t)$ .  
Right panels show aggregate loss distributions averaged over time, labeled with corresponding statistics of interest.  
\textbf{B)} Depictions of the range of the joint distributions of $(R_1,R_2)$ (dark blue) and $(Q_1,Q_2)$ 
(light blue).  The covariance and correlation are 
important for calculating aggregate capital different risk categories, but can depend on time window of observations.
}
\end{figure}

Our main focus is on frequency distribution modeling because our aim is to show how inappropriately processing 
data in time can lead to unexpected results.  Indeed, a simple example in Fig \ref{fig1:setup} shows how the time 
window of cumulative losses can dramatically alter the loss distribution statistics.  The convention we use throughout is 
that the \underline{dark blue} denotes losses in a small time window $\Delta t$, \underline{light blue} denotes losses in a larger time window $T_w$.  
The actual loss time series is \underline{$R$}, and cumulative losses in window $T_w$ is \underline{$Q$}.

The outline of the paper is as follows:
\begin{itemize}
    \item Describe and analyze the Inhomogeneous Poisson Process frequency distribution model for single loss time series. 
    \item Develop theory for capturing statistics of cumulative losses in arbitrary time windows.
    \item Extend Inhomogeneous Poisson Process model to multiple loss time series, allowing input correlations in frequency distribution and heterogeneity across risk categories. 
    Derive formulas for covariance of loss distributions in arbitrarily large time windows. 
    \item Apply theory to data averages obtained from consortium to demonstrate why careful consideration of time windows for loss reporting is necessary.
\end{itemize}

\noindent
\textbf{\underline{Background assumptions}}

The loss distribution model consists of both a frequency and a severity distribution for a given operational risk loss categories \citep{bis_2006,chern08book}. To enable mathematical analysis, 
we make two common assumptions: the time series is stationary 
(i.e., the statistics do not vary over time), and that frequency is independent of severity \citep{bis_2006,chern08book}. That is:
\begin{equation}
    \mu(t_0)=\mu(t_1)=\mu_X.
\end{equation}
We let $I(t)$, the frequency distribution, indicate whether a loss event occurred at time $t$ 
\begin{equation}
I(t) =
\begin{cases} 
     \ 0, & \text{no loss event at time $t$} \\
      \ 1, & \text{have loss event at time $t$}.\\
\end{cases}
\end{equation}
Let $S$ denote the random size of the loss (severity).  The loss time series $R(t)$ is:
\begin{equation}\label{eqn:R_homo}
	R(t) = I(t) * S
\end{equation}
The independence assumption is:
\begin{equation}
	f_{I , S } = f_I * f_S 
\end{equation}
where $f_X$ denotes the marginal distribution (PDF) of $X\in\{I,S\}$.

Note that time $t$ is theoretically continuous, loss events occur at discrete time points as a point process, and loss magnitudes $S$ are continuous.  
However, since the model is represented on a computer, we use discrete time $\left( \Delta t \right)$. 
The resulting formulas presented here account for time step length $\Delta t$ to make a direct correspondence with our publicly available computer code, and 
should thus be more useful to other practitioners.

\subsection{Time series with inhomogeneous Poisson Process frequency distribution}\label{sec:Inhomog}

Realistic models of loss frequencies must account for (time-)dependence between loss events; 
operations of institutions are not simply memory-less like an homogenous Poisson Process.  
Here we include time-varying dynamics via an 
inhomogeneous Poisson Process model, where the probability of $R(t)>0$ depends on time.  
Throughout the rest of the paper, we will refer to standard methods used to calculate loss statistics for a simple 
homogeneous Poisson Process model for the frequency distribution, which are provided in Appendix \ref{sec:homog}A. 
Note that the resulting loss time series when the frequency distribution is a Poisson Process is commonly referred to as a \underline{marked Poisson Process} 
\citep{pplect17} where the mark is a continuous positive random variable corresponding to the severity. 

The direct calculation of the (co-)variance of cumulative losses in larger windows $T_w$ is unwieldy because of the 
large number (${n \choose 2}$ where $n$ is the number of observations in $T_w$) 
of different $R_j R_k$ correlated terms, and the distribution of the number of loss events is generally intractable with temporal correlations.  
We use an approximation to relate the autocovariance $A(t)$ to the variance of cumulative losses in time window $T_w$ \citep{kay1993,shea2008correlation,litwin2011balanced,litwin2012spatial,BarreiroLy_RecrCorr_17,BarreiroLy_jmns_18}: 
\begin{equation}\label{eqn:autVar}
    \sigma^2_{Q} = \int_{-T_w}^{T_w} A(t)\Big( T_w - \vert t \vert \Big)\,dt
\end{equation} 
because the autocovaraince $A(t)$ is tractable (this formula assumes $A(t)$ is sufficient for calculating $ \sigma^2_{Q}$, see Materials and methods section for details).  

Here $R(t) = S * I(t)$, with $P(I(t)=1)=\nu(t)\Delta t$ where the time points $t$ are spaced a part by $\Delta t$, 
and where $\nu(t)$ is the probability per unit time of a loss event occurring.  We assume $\nu(t)$ is governed by a stochastic 
differential equation:
\begin{equation}\label{eqn:inhomogFreq}
 \tau \frac{d}{dt}\nu(t)=-\nu(t) + \tau a \sum_{k} \delta(t-t_k)   
\end{equation}
where $t_k$ are random points drawn from a homogeneous Poisson Process with rate $\gamma$, 
$a$ is the jump size of $\nu(t)$ at times $t_k$, and $\tau$ is the time-scale that determines how fast $\nu(t)$ decays to 0 
in the absence of random jumps at $t_k$.  The 
dynamics are essentially determined by two main factors: $(\gamma,a)$ controls the magnitude of increases in frequency, 
and $\tau$ is a measure of the memory (larger values correspond to longer memory).  In contrast to some other time series models, this 
stochastic differential equation formulation consists of parameters that are easy to understand: magnitude and memory or time-scale. 
This frequency model (Eq.\eqref{eqn:inhomogFreq}) is a simple generalization of a homogeneous Poisson Process, 
seen by taking the limit $\tau\to0$, $a\to\infty$ so that $a\tau=1$. This model has not been precisely calibrated to data (but see application to data below). 
The model is very similar in form to a Hawkes process \citep{hawkes71,hawkes18} commonly used in finance and numerous 
applications; it is a point process model where the probability rate of an event (randomly) changes at prescribed points and could be followed by exponential decay. 

\begin{figure}[!htb]
\centering
\includegraphics[width=.75\columnwidth]{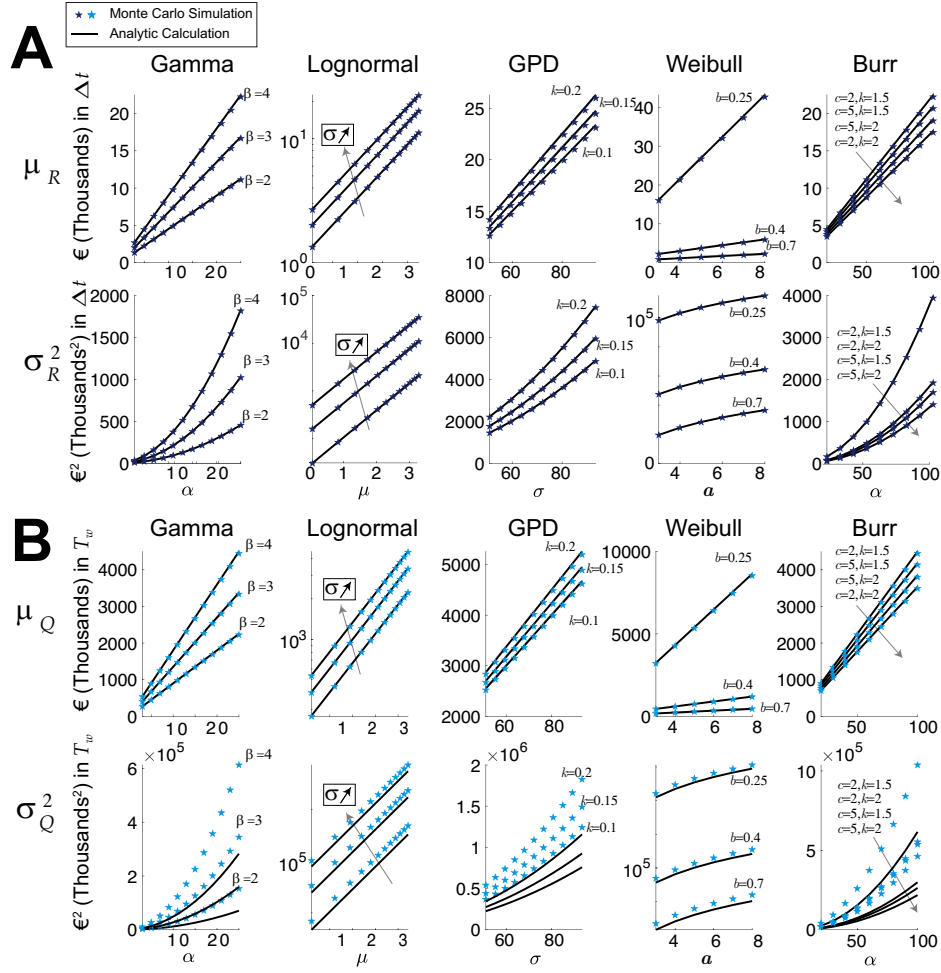}
\caption{\label{fig:hetCompare} \textbf{Validation of results for inhomogeneous Poisson Process frequency model \eqref{eqn:inhomogFreq}.} 
Comparing mean and variance of aggregate loss time series with inhomogeneous Poisson Process 
frequency distribution (here we set $\tau=1.2, a=1, \gamma=75/2$) for 
common severity distributions (see Table \ref{tbl:stats}).  Note that Table \ref{tbl:riskAvgs} shows that 
this average frequency is typical in industry-wide reporting.  
\textbf{A}) Comparing the analytic formula for $\mu_R$ (Eq.\eqref{eqn:mnRinhomog}, 
$\sigma^2_R$ (Eq.\eqref{eqn:inhomVarR}) (solid lines) 
with the Monte Carlo simulations (stars), and similarly for $Q$ in (\textbf{B}): $\mu_Q$ (Eq.\eqref{eqn:mnSever_inhom}), 
and $\sigma^2_Q$ (Eq.\eqref{eqn:inhomVarRsc}), with $\Delta t=0.001$ in time units of years ($\approx 0.365$ day).  
For Lognormal, $\sigma=\sqrt{2\log(2)}, \sqrt{2\log(3)}, 
\sqrt{2\log(4)}$; ; vertical axis is log-scale for Lognormal, and Weibull variance.  
The results for $\sigma^2_Q$ are not 
as accurate, but the general trends are captured (see Fig. \ref{fig:ScaledSigm}A)
The Monte Carlo simulations are for 50,000 realizations with each times series 100 years long. 
The severity distribution parameters vary greatly because we 
perform a large parameter sweep, but the average is about 10 to 100 \euro{} per loss event, 
so that average yearly loss is on the order of 10s of millions of \euro{}, consistent with industry reporting 
(Table \ref{tbl:riskAvgs}).   
}
\end{figure}

We are interested in the statistics of the loss time series $R_j(t)$ and $Q_j(t)$, including the mean $\mu_R$ and variance $\sigma^2_R$ in both small time windows 
and arbitrarily large windows $T_w$: $\mu_{Q}$, $\sigma^2_{Q}$.  
Thus we derive analytic formulas for these entities (see Materials and methods section (Eq \eqref{eqn:mnRinhomog}, \eqref{eqn:inhomVarR}, \eqref{eqn:mnSever_inhom}, \eqref{eqn:inhomVarRsc}; 
black lines in Fig \ref{fig:hetCompare})); we also state the corresponding entities for the frequency distribution ($\nu(t)$ and $\mathcal{V}$ for sum 
of events in $T_w$) that will be instrumental in fitting our model to data:
\begin{eqnarray}
    \mathbb{E}[\nu(t)] &=& a\tau \gamma \nonumber \\
    Var(\nu(t)) &=& \frac{a^2\gamma \tau}{2} \nonumber \\
    \mathbb{E}\left[\mathcal{V}\right]  &=&  T_w a\gamma\tau  \nonumber \\
    Var\left(\mathcal{V}\right) &=& a^2\gamma \tau^2 \Big( T_w + \tau(e^{-T_w/\tau} -1 ) \Big) \nonumber \\
    \mu_{R} &=& \mu_S(a\gamma \tau \Delta t) \nonumber \\
    \sigma^2_{R} &=& \mu_{S^2}(a\tau\gamma\Delta t) - (\mu_{S}a\tau\gamma\Delta t)^2 \nonumber \\
    \mu_{Q}  &=&  T_w a\gamma\tau \mu_S \nonumber \\
    \sigma^2_{Q} &=& 2\frac{\sigma^2_{R}}{\Delta t} \tau \Big( T_w + \tau(e^{-T_w/\tau} -1 ) \Big) \nonumber
\end{eqnarray}
These formulas are generally very accurate in comparison to the Monte Carlo simulations (stars, Fig \ref{fig:hetCompare}). 
The glaring exception is the analytic approximation of $\sigma^2_{Q}$ in Fig \ref{fig:hetCompare}B (bottom row, black lines); 
our theory does not accurately capture the Monte Carlo simulations (light blue stars).  

\underline{\textbf{Explanation of $\sigma^2_{Q}$ mismatch:}} This was initially surprising to us, previously believing that there might have been an error in our analytic 
calculation and/or in the Monte Carlo simulations.  After much investigation and careful re-checking, 
we are certain that this quantitative mismatch is not easily rectifiable in these operational risk models. 
Indeed, others have found there can be discrepancies with simulations and this same calculation (Eq. \eqref{eqn:inhomVarRsc}) 
to approximate statistics in different time windows.  
Although the details of the models are different, we 
see this in the variance/covariance of (spike) statistics in many other contexts 
(e.g., see Fig 2C, Fig 5B, Fig 6B2 in \citet{litwin2012spatial}, Fig 3B,C,D in \citet{litwin2011balanced}, Fig 2 in \citet{BarreiroLy_RecrCorr_17}); 
the important trends are captured by the formula but there is not precise quantitative matches to Monte Carlo simulations. 

\begin{figure}[!htb]
\centering
\includegraphics[width=\columnwidth]{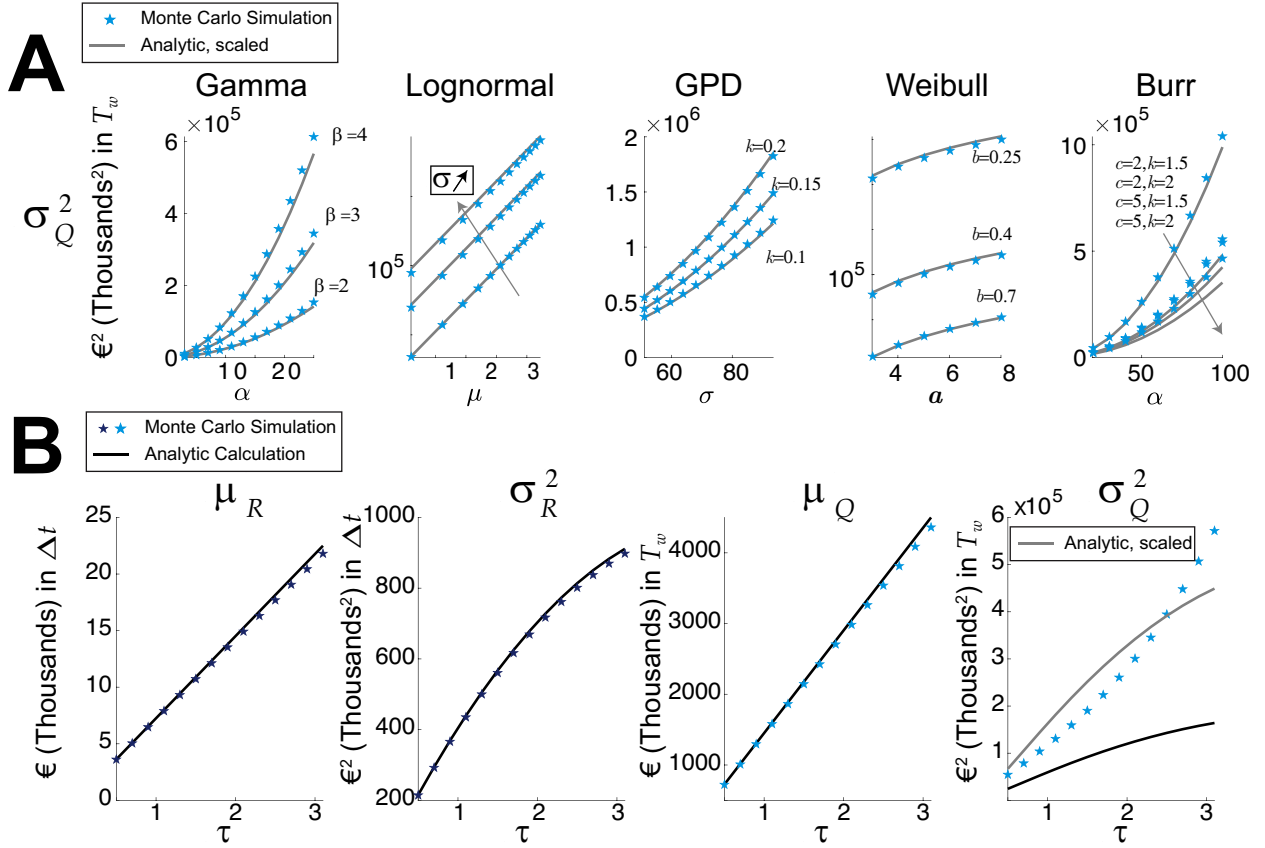}
\caption{\label{fig:ScaledSigm} \textbf{ A simple scale factor can yield accurate results.}  
Quantitative inaccuracies between the Monte Carlo 
simulation and the analytic formula (Eq.\eqref{eqn:inhomVarRsc}) for $\sigma^2_Q$ are fixed with a single scaling factor. 
\textbf{A}) Same as Figure \ref{fig:hetCompare}B, except analytic formula are scaled by 2, 1.5, 1.6, 1.5, 1.6, respectively. 
The scaling factors were manually determined. 
\textbf{B)} Fixing all parameters ($a=1, \gamma=75/3.1$) except $\tau$ that varies from 0.5 to 3.1 (unit=years), 
using a Gamma distributed severity distribution $S\sim$Gamma$(\alpha=20,\beta=3)$ ; last column has $\sigma^2_Q$ scaled by 2.74. The accuracy of the theory slightly diminishes as $\tau$ increases. 
}
\end{figure}

Our analytic calculation still has value in capturing the qualitative trends of the Monte Carlo simulations.  A simple scalar 
factor can yield relatively accurate approximations (Fig \ref{fig:ScaledSigm}), indicating that our formula 
\begin{equation}\label{eqn:inhVR2}
    \sigma^2_{Q}  = 
    2\frac{\mu_{S^2}(a  \tau \gamma \Delta t) - \left(\mu_S a \tau \gamma \Delta t\right)^2}{\Delta t} \tau \Big( T + \tau(e^{-T/\tau} -1 ) \Big)
\end{equation}
reveals the relative trends as parameters are varied. Fig \ref{fig:ScaledSigm}A shows the comparisons of the analytic theory (black curve) 
but also with a scaling factor (gray curves); the parameters here are the same same parameters as bottom row in Fig \ref{fig:hetCompare}B).  
Fig \ref{fig:ScaledSigm}B shows how our analytic calculation approximates the statistics as the time-scale of the inhomogeneous 
frequency distribution, $\tau$, varies; as $\tau$ increases there appears to be 
more discrepancies between the Monte Carlo and the analytic calculations. 
The last panel on the far-right for $\sigma^2_{Q}$ shows both the 
original formula (black) and the curve scaled by 2.74 (gray; computed as least-squares fit to the stars). 

\begin{figure}[!htb]
\centering
\includegraphics[width=\textwidth]{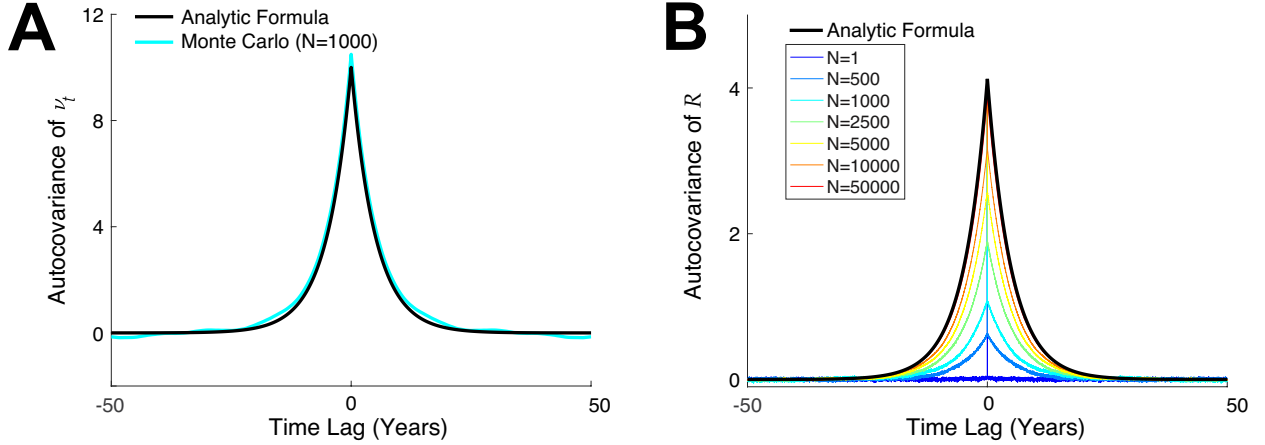}
\caption{\label{fig:varyN} \textbf{ The number of realizations effects Monte Carlo of $A_R(t)$ significantly.} 
Consider the inhomogeneous Poisson Process model for the frequency distribution in \eqref{eqn:nut2} 
($\gamma=4$ events/year, $a=1$ event/year, $\tau=5$ years), 
with a gamma distribution for the severity $S\sim$Gamma$(\alpha=1.5,\beta=2.5)$. 
\textbf{A)} Comparison of the auto-covariance of the frequency distribution: $A_{\nu(t)}(t)$, calculated from Monte Carlo simulations (cyan) of Eq. \eqref{eqn:inhomogFreq} 
with theoretical calculation (black, from Eq. \eqref{eqn:nut2}) shows good agreement with relatively few realizations ($N=1000$, 1000 years for each realization). 
\textbf{B)} Comparisons of $A_{R}(t) \Delta t$ calculated from Monte Carlo simulations (rainbow) with 
theoretical calculation (black, from Eq. \eqref{eqn:autcR}) shows that many realizations are required to achieve accuracy. 
}
\end{figure}

\underline{\textbf{Technical remark:}} in Figures \ref{fig:hetCompare}--\ref{fig:ScaledSigm}, rather than simulating a realization of 
a long time series (as in Fig. \ref{fig:homoCompare}), we simulated many realizations (50,000) of a moderate length time. This was to insure that 
the autocovariance of $R(t)$, $A_R(t)$ was accurately captured compared to the theory (Eq. \eqref{eqn:autcR}); see 
Figure \ref{fig:varyN}B with 50,000 realizations in red.  
Note that a very small number of realizations can accurately capture the 
autocovariance of the time-varying inhomogeneous Poisson rate $\nu(t)$ (Fig. \ref{fig:varyN}A).  

\subsection{Multiple loss time series}\label{sec:multloss}

We extend the methods for a single loss category to two loss categories, which often suffices for calculating aggregate capital because pairwise correlations are commonly used to determine 
diversification of risk \citep{chern08book}.  
Although we assume that the frequency distributions for both loss 
categories have the same form as Eq. \eqref{eqn:inhomogFreq}: 
\begin{eqnarray}\label{eqn:inhFreq2}
 \tau_1 \nu_1' & =& -\nu_1 + \tau_1 a_1 \sum_{k_1} \delta(t-t_{k_1})  \\
 \tau_2 \nu_2' & =& -\nu_2 + \tau_2 a_2 \sum_{k_2} \delta(t-t_{k_2})  \label{eqn:inhFreq2b}
\end{eqnarray}
a key difference is that the frequencies can be correlated. 
The random times $t_{k_1}, t_{k_2}$ are again governed by an inhomogeneous Poisson Process with rates $\gamma_1$ and $\gamma_2$, respectively, 
but the random times $(t_{k_1},t_{k_2})$ can be correlated.  The parameter $c\in[-1,1]$ is a measure of 
the correlation between $t_{k_1}, t_{k_2}$; letting $\bar{\gamma}:=\min(\gamma_1,\gamma_2)$, the value $\vert c\vert \bar{\gamma}$ is the probability per unit time 
that $\nu_1$ and $\nu_2$ instantaneously jump at the same time, both in the positive direction if $c>0$ and in opposite directions 
when $c<0$.  
The marginal statistics of $\nu_j$, $\mathcal{V}_j$, $R_j$ and $Q_j$, for $j\in\{1,2\}$, are the same as with a single loss time series:
\begin{eqnarray}
    \mathbb{E}[\nu_j] &=& a_j\tau_j \gamma_j \\
    Var(\nu_j) &=& \frac{a_j^2\gamma_j \tau_j}{2} \\
    \mathbb{E}\left[\mathcal{V}_j\right]  &=&  T_w a_j\gamma_j\tau_j   \\
    Var\left(\mathcal{V}_j\right) &=& a_j^2\gamma_j \tau_j^2 \Big( T_w + \tau_j(e^{-T_w/\tau_j} -1 ) \Big) \\
    \mu_{R_j} &=& \mu_{S_j}(a_j\gamma_j \tau_j \Delta t)  \\
    \sigma^2_{R_j} &=& \mu_{S_j^2}(a_j\tau_j\gamma_j\Delta t) - (\mu_{S_j}a_j\tau_j\gamma_j\Delta t)^2 \\
    \mu_{Q_j}  &=&  T_w a_j\gamma_j\tau_j \mu_{S_j}  \\
    \sigma^2_{Q_j} &=& 2\frac{\sigma^2_{R_j}}{\Delta t} \tau_j \Big( T_w + \tau_j(e^{-T_w/\tau_j} -1 ) \Big).
\end{eqnarray}

\begin{figure}[!htb]
\centering
\includegraphics[width=\textwidth]{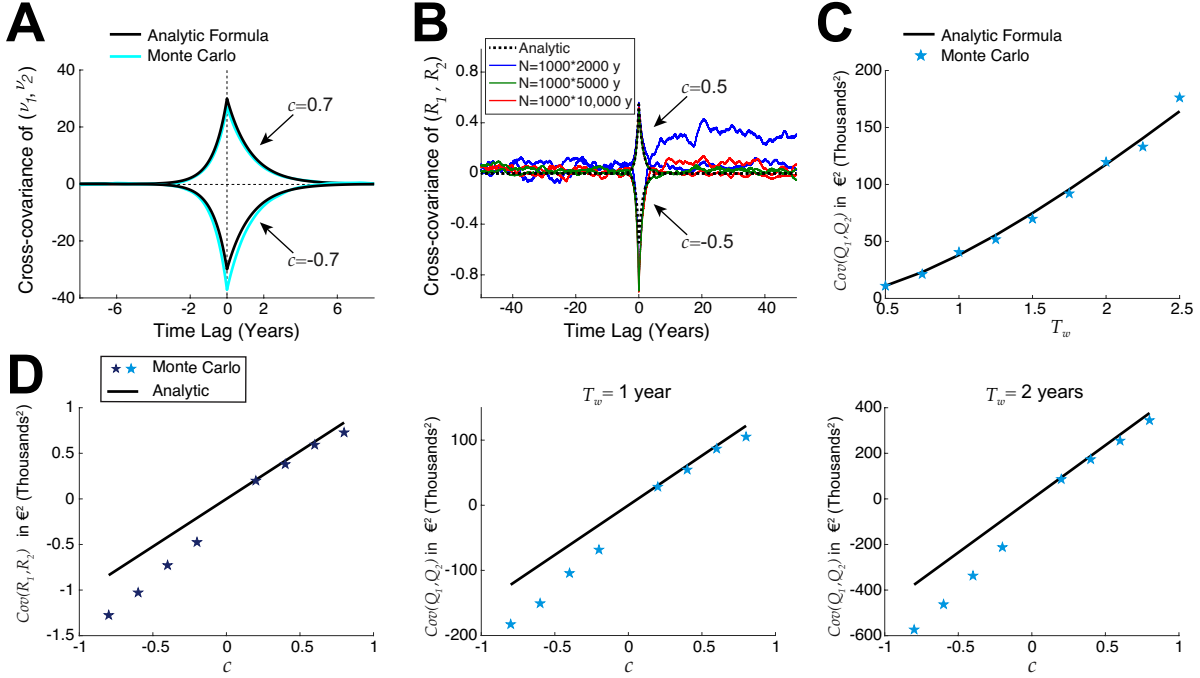}
\caption{ \label{fig:cov1} \textbf{ Analytic theory capture covariance of cumulative losses in different time windows.} 
\textbf{A}) Comparisons of cross-covariance of frequency rates $(\nu_1,\nu_2), CC_{\nu}(t)$; 
Monte Carlo (cyan) with very few realizations and analytic theory in solid black (Eq \eqref{eqn:crssnu}) 
with input correlations $c=\pm0.7$. 
\textbf{B}) Comparisons of cross-covariance function of actual loss time series $CC_{R}(t)$; 
Monte Carlo in colors with fixed $N=1000$ but total time 2000 years (blue), 
5000 years (red), 10,000 years (green) and analytic theory (black, Eq \eqref{eqn:covTw1}), 
with input correlations $c=\pm0.5$. 
\textbf{C}) Comparisons for $Cov(R_1,R_2)$ and $Cov(Q_1,Q_2)$ 
with $T_w=1$ year (middle) and $T_w=2$ years (right), same format as before. The input correlation $c$ 
varies across a wide-range. 
\textbf{D}) Varying time windows $T_w$ for a fixed $c=0.25$. 
The fixed parameters are: $a_1=1.5, \tau_1=1.3\,\text{years}, \gamma_1=30\,\text{years}^{-1}, 
a_2=2, \tau_2=0.75\,\text{years}, \gamma_2=40\,\text{years}^{-1}$; $S_1\sim$GPD$(k=.15,\sigma=50)$, 
$S_2\sim$Weibull$(a=5,b=0.4)$.
}
\end{figure}

\begin{figure}[!htb]
\centering
\includegraphics[width=\textwidth]{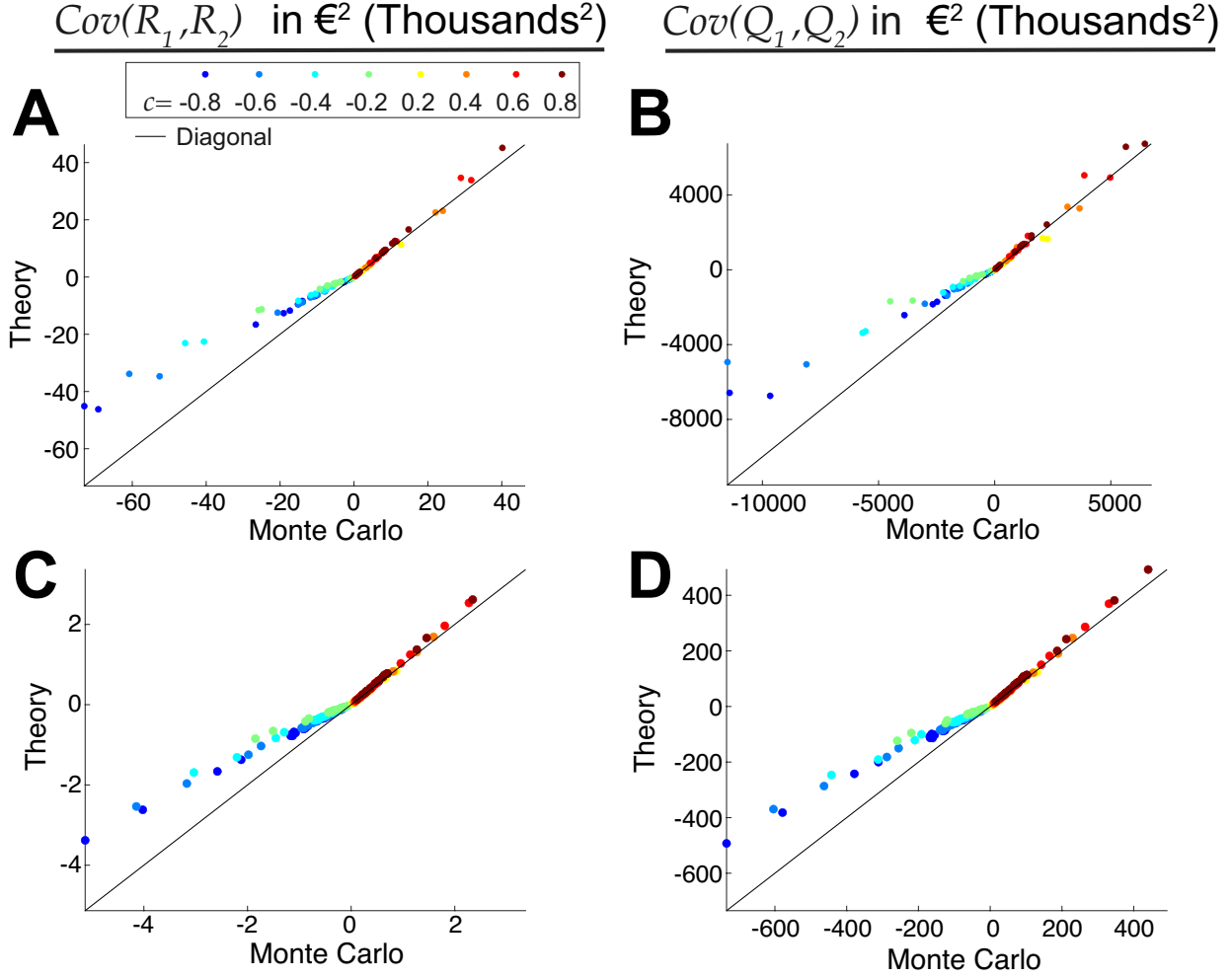}
\caption{\label{fig:covRand} \textbf{Analytic theory for covariance of cumulative losses is accurate for random parameters.} 
Here we set $T_w=1$\,year in \textbf{B,D}, but we vary the input correlation $c$ (see legend for coloring), with 20 
randomly chosen severity distribution parameters for each $c$; severity parameters are all uniform distributions (independent) with 
the same ranges as in Figs \ref{fig:hetCompare},\ref{fig:homoCompare}.  The 
horizontal axes is the Monte Carlo, vertical is the analytic theory; the diagonal line is solid 
black, so perfect accuracy are points that are on the black line. 
\textbf{A})--\textbf{B}) 
$S_1\sim$Lognormal, $S_2\sim$GPD, with $\Delta t$ windows (\textbf{A}) and $T_w=1$\,year (\textbf{B}). 
\textbf{C})--\textbf{D}) 
$S_1\sim$Weibull, $S_2\sim$Burr, with $\Delta t$ windows (\textbf{C}) and $T_w=1$\,year (\textbf{D}).  
The fixed parameters are: $a_1=1.5, \tau_1=1.3\,\text{years}, \gamma_1=30\,\text{years}^{-1}, 
a_2=2, \tau_2=0.75\,\text{years}, \gamma_2=40\,\text{years}^{-1}$; 
}
\end{figure}

In the Materials and methods section, we derive an equation for $Cov(Q_1,Q_2)$ that 
is based on:
\begin{equation}\label{eqn:cov_wind}
    Cov(Q_1,Q_2) = \int_{-T_w}^{T_w} CC_R(t)\Big( T_w-\vert t \vert \Big)\,dt.
\end{equation}
See Eq \eqref{eqn:covTw1}; the result is:
$$ Cov(Q_1,Q_2)=c\bar{\gamma} \mu_{S_1}\mu_{S_2}  a_1 a_2\frac{\tau_1\tau_2}{\tau_1+\tau_2} \Big[ \tau_1\Big(T_w+\tau_1(e^{-T_w/\tau_1}-1)\Big) + \tau_2\Big(T_w+\tau_2(e^{-T_w/\tau_2}-1)\Big) \Big]\Delta t. $$
Figs \ref{fig:cov1}--\ref{fig:covRand} shows a summary of the components of our analytic theory, which overall 
is accurate. In all panels of Fig \ref{fig:cov1}, the 
two severity distributions are fixed with $S_1\sim$GPD$(k=.15,\sigma=50)$, 
$S_2\sim$Weibull$(a=5,b=0.4)$. 
Fig\ref{fig:cov1}A shows comparisons of the cross-covariance of the rates $(\nu_1,\nu_2)$ 
of the 2 different frequency distributions: $CC_{\nu}(t)$ (cyan is Monte Carlo, black is 
calculation Eq \eqref{eqn:crssnu}). Here we show two input correlations $c=\pm0.7$ (see legend 
for parameters of Eqs \eqref{eqn:inhFreq2}--\eqref{eqn:inhFreq2b}). The analytic theory tends to 
underestimate the cross-covariance of the Monte Carlo, a consistent trend that holds for all parameters 
we considered (see below). 
Comparisons of cross-covariance of the actual loss time series $(R_1,R_2)$ unsurprisingly shows 
that more realizations matches the analytic theory (Eq \eqref{eqn:covTw1}, black dashed) better; that is, 
the red curve is closest to the black-dashed curves than all of the other colors in 
Fig \ref{fig:cov1}B (two input correlations $c=\pm0.5$). 
Fig \ref{fig:cov1}C show the accuracy of our analytical equation for different time windows $T_w$ for a fixed $c=0.25$. 
Finally, Fig \ref{fig:cov1}D shows comparisons for $Cov(R_1,R_2)$ and $Cov(Q_1,Q_2)$ 
with $T_w=1$ year and $T_w=2$ years, varying the input correlation $c$ over a large range of values. 
The theory performs much better with positive input correlation than negative.

Fig \ref{fig:covRand} has more demonstrations of the 
analytic theory for covariance of cumulative losses. 
We vary the input correlation $c$ (see legend for coloring) and 
randomly choose the severity distribution parameters, which are all independent 
uniform distributions with 
the same ranges as in Fig \ref{fig:hetCompare} (and Fig \ref{fig:homoCompare} in Appendix A). The 
horizontal axes is the Monte Carlo, vertical is the analytic theory; the diagonal line is solid black, so perfect accuracy are points that lie on the diagonal. 
A--B shows severity: $S_1\sim$Lognormal, $S_2\sim$GPD; 
C--D shows severity: $S_1\sim$Weibull, $S_2\sim$Burr, where 
$\Delta t$ windows are shown in A,C and $T_w=1$\,year is shown in B,D. Once 
again, positive input correlations (redish) result in better matches than negative (bluish).

Recall that the (point-wise) covariance of the frequency distributions is:
\begin{equation}\label{eqn:covFreq}
    Cov(\nu_1,\nu_2)=c\bar{\gamma}a_1a_2\frac{\tau_1\tau_2}{\tau_1+\tau_2}. 
\end{equation}

Note that $Cov(\nu_1,\nu_2)$ in larger time windows $T_w$ is:
\begin{equation}\label{eqn:covFreqs_T}
  c\bar{\gamma}  a_1 a_2\frac{\tau_1\tau_2}{\tau_1+\tau_2} \Big[ \tau_1\Big(T_w+\tau_1(e^{-T_w/\tau_1}-1)\Big) +\tau_2\Big(T_w+\tau_2(e^{-T_w/\tau_2}-1)\Big) \Big]. 
\end{equation}
which looks very similar to the formula for $Cov(Q_1,Q_2)$ -- this is not surprising since these equations were derived from integrating the 
autocovariances of $A_{\nu}$ or $A_R$ that only differ by a scaling factor. 
We state this equation for completeness, and also because it will be used in the next section where we apply our theory to real data.

\subsection{Applying our model to industry-wide averages}\label{sec:applyTheory}

\begin{figure}[!htb]
\centering
    \includegraphics[width=\textwidth]{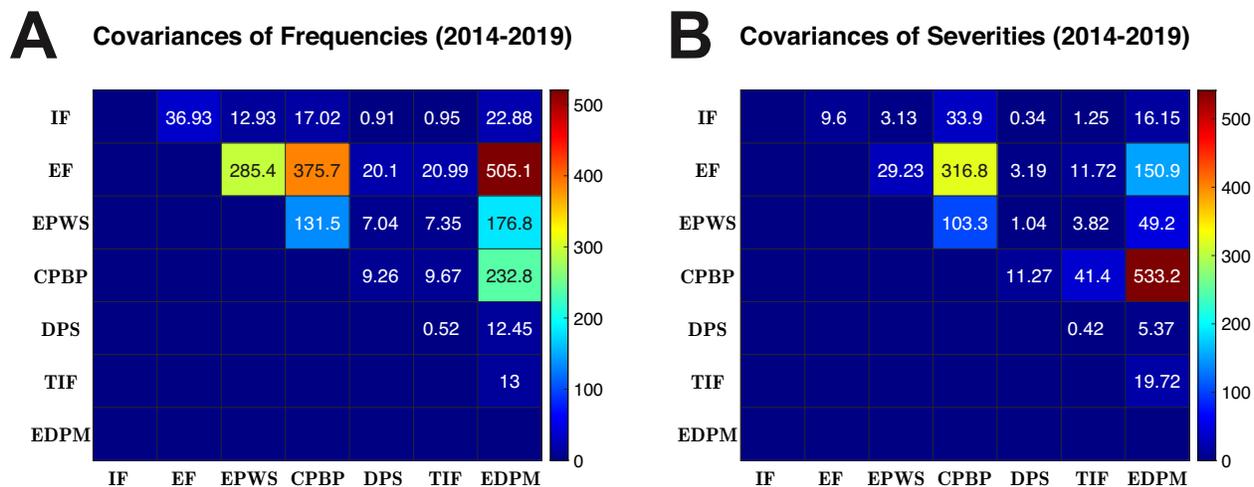}
    \caption{\label{fig:covORX} \textbf{The year-to-year covariances of ORX data}. 
\textbf{A}) The covariances of the average frequencies per institution by risk category (see Table \ref{tbl:riskAbbrev} for 
definition of abbreviations). 
\textbf{B}) Same as \textbf{A} but for average severity per institution (in \euro{}-Millions).  
\textbf{Note that (B) is not used in the model fitting}, we provide this for completeness. 
Excluding diagonal (variances) and lower triangular portion because of symmetry, see Table \ref{tbl:riskAvgs} for the univariate statistics by risk categories. 
}
\end{figure}

Obtaining actual operational risk loss data with granular details for a particular institution is extremely difficult due to proprietary 
reasons.  Loss information could reveal vulnerabilities in operations that institutions do not want to publicize, 
especially to business competitors.  The most easily attainable and detailed operational risk loss 
data we found was provided by \textbf{ORX}, an organization that facilitates sharing of actual operational risk losses among 
its member institutions in a secure and anonymous platform \citep{orx_url}.  
ORX provides information to the public about the total count of loss events and the amount of losses 
in prior years, as well 
as the number of member institutions contributing data in a given year \citep{orx_urlData1,orx_urlData2}.  
The cumulative frequency and severity of losses are available by risk categories, see Table \ref{tbl:riskAvgs}.  
We also use the year-to-year covariance of the frequency of losses by risk category shown in Fig \ref{fig:covORX}A.  
For completeness, we show the year-to-year severity of losses in Fig \ref{fig:covORX}B, but this will not be used (see explanation below).  
Details on how the ORX data was obtained are outlined in the Materials and methods section. 

Important issues to keep in mind: this data provided by ORX is a coarse industry-wide 
average with contributions from many different institutions, and does not contain the precise times or magnitudes of individual loss events. Thus, the application of our 
method here is a demonstration, i.e., a proof of principle. With the AMA, external coarse data from ORX is used to merely supplement capital calculations; institutions likely rely on their own internal proprietary data more heavily in their model fitting. 

\begin{table}[!htb]
\centering
\caption{\textbf{Risk Categories and Abbreviation} 
The convention used for Operational Risk loss categories are 
segmented into the 7 categories below. Although the names of each 
category can vary slightly, the actual descriptions are equivalent; we adopt the naming convention used in ORX \citep{orx_urlData1,orx_urlData2}.}
\begin{tabular}{ll}
\multicolumn{2}{l}{\textbf{Risk Categories}}                                            \\ \thickhline
\multicolumn{1}{l|}{\textbf{Abbreviation}} & \textbf{Definition}                        \\ \thickhline
\multicolumn{1}{l|}{\textbf{IF}}           & Internal Fraud                             \\
\multicolumn{1}{l|}{\textbf{EF}}           & External Fraud                             \\
\multicolumn{1}{l|}{\textbf{EPWS}}         & Employment Practices, Workplace Safety     \\
\multicolumn{1}{l|}{\textbf{CPBP}}         & Clients, Products, Business Practices      \\
\multicolumn{1}{l|}{\textbf{DPS}}          & Disasters and Public Safety                \\
\multicolumn{1}{l|}{\textbf{TIF}}          & Technology and Infrastructure Failure      \\
\multicolumn{1}{l|}{\textbf{EDPM}}         & Execution, Delivery and Process Management
\end{tabular}
\label{tbl:riskAbbrev}
\end{table}

\begin{table}[!htb]
\centering
\caption{\label{tbl:riskAvgs} \textbf{Statistics by Risk Categories.} 
Columns show: the average frequency (\# events in a year, per institution) and 
the variances (absent from Fig. \ref{fig:covORX}A), segmented by the 7 risk 
categories.  The last column is the severity (\euro-Millions) \textbf{per event}, 
which is the only information we have access to for fitting our models to $\mu_S$.  
The overall average frequency and severity without regard to category are: 514.13 events/year and 321.68 \euro-Millions/event  (resp.).  
Same ORX data the same source as in Figure \ref{fig:covORX}.}
\resizebox{\textwidth}{!}{
\begin{tabular}{l|cc||c}
\multicolumn{1}{c|}{\multirow{3}{*}{\textbf{Risk Category}}} & \multicolumn{2}{c|}{\textbf{Statistics (over 6 years)}}                                            & \textbf{Grand Average}                                            \\ \cline{2-4} 
\multicolumn{1}{c|}{}                                        & \multicolumn{1}{c|}{\textbf{Frequency Mean}}       & \textbf{Frequency Var}                        & \textbf{Severity}                                                 \\
\multicolumn{1}{c|}{}                                        & \multicolumn{1}{c|}{(\# events per year)} & \multicolumn{1}{l|}{(\# events per year)$^2$} & \multicolumn{1}{l}{(\euro{}-Millions per event)} \\ \hline
\textbf{IF}                                                  & \multicolumn{1}{c|}{9.22}                          & 1.67                                          & $250.91$                                                          \\
\textbf{EF}                                                  & \multicolumn{1}{c|}{203.51}                        & 815.19                                        & $106.23$                                                          \\
\textbf{EPWS}                                                & \multicolumn{1}{c|}{71.25}                         & 99.91                                         & $98.95$                                                           \\
\textbf{CPBP}                                                & \multicolumn{1}{c|}{93.8}                          & 173.16                                        & $814.51$                                                          \\
\textbf{DPS}                                                 & \multicolumn{1}{c|}{5.02}                          & 0.496                                         & $153.3$                                                           \\
\textbf{TIF}                                                 & \multicolumn{1}{c|}{5.24}                          & 0.54                                          & $539.31$                                                          \\
\textbf{EDPM}                                                & \multicolumn{1}{c|}{1.26}                          & 312.93                                        & $288.58$                                                         
\end{tabular}
}
\end{table}

\subsubsection{Fitting model parameters to ORX data}

Although ORX data has some yearly statistics of the severity of losses by risk categories, there is not enough granular detail to estimate the parameters of a 
given severity distribution.  We can only use the ORX data to specify the mean severity \textit{per event} in a given risk category $j$: $\mu_{S_j}$, the higher order statistics (e.g., variance of 
loss \textit{per event}) are not accessible.  Fortunately, the formula we derived for the covariance of losses in $T_w$ \underline{only} 
relies on the mean severities of each event $\mu_{S_{j/k}}$ (via Eq \eqref{eqn:covTw1}):  

$$ Cov(Q_j,Q_k)=c_{j,k}\bar{\gamma} \mu_{S_j}\mu_{S_k}  a_j a_k\frac{\tau_j\tau_k}{\tau_j+\tau_k} \Big[ \tau_j\Big(T_w+\tau_j(e^{-T_w/\tau_j}-1)\Big) + \tau_k\Big(T_w+\tau_k(e^{-T_w/\tau_k}-1)\Big) \Big]\Delta t. $$

So fitting our model to ORX data mainly involves the frequency distribution, i.e., fitting $c_{j,k}  (j\neq k), a_j,\tau_j, \gamma_j$, giving a total of 42 parameters to determine with the 7 loss categories 
(7 different $a,\tau,\gamma$ and 21 input correlations $c_{j,k} (j\neq k)$).  We naturally want the mean frequency from the model $\mu_{\nu_j}=a_j\tau_j\gamma_jT_w$ to equal the yearly frequency average from ORX data (Table \ref{tbl:riskAvgs}, 1st column), and similarly for the variances (Table \ref{tbl:riskAvgs}, 2nd column) and covariances (Fig \ref{fig:covORX}A).  
Thus our objective is to fit the parameters to the following system:

\begin{eqnarray}
	a_j \tau_j \gamma_j T_w &=& \mu_{\nu_j,ORX}   \text{   ; for }j=1,\dots,7  \nonumber \\
	a_j^2\gamma_j \tau_j^2 \Big( T_w + \tau_j(e^{-T_w/\tau_j} -1 ) \Big) &=& \sigma^2_{\nu_j,ORX}    \text{   ; for }j=1,\dots,7  	\nonumber 
\end{eqnarray}
\begin{eqnarray}	
	c_{j,k}\bar{\gamma}  a_j a_k\frac{\tau_j\tau_k}{\tau_j+\tau_k} \Big[ \tau_j\Big(T_w+\tau_j(e^{-T_w/\tau_j}-1)\Big) +\tau_k\Big(T_w+\tau_k(e^{-T_w/\tau_k}-1)\Big) \Big] =   \nonumber \\
	 Cov(\nu_{j,ORX},\nu_{k,ORX}) \text{   ; for }j\neq k \nonumber
\end{eqnarray}
Since the ORX data provides yearly averages, we set $T_w=1$\,year.

Summary of model parameter fitting procedure:
\begin{itemize}
\item We set $\mu_{S_j}$ equal to the corresponding value in Table \ref{tbl:riskAvgs}, right-most column. 
\item Find 42 parameters ($c_{j,k}, a_j,\tau_j,\gamma_j$) simultaneously using 35 equations from ORX data (7 mean and variances of 
yearly frequency in first 2 columns of Table \ref{tbl:riskAvgs}, 21 covariance frequencies in Fig \ref{fig:covORX}A). 
\end{itemize}

\begin{figure}[!htb]
\centering
    \includegraphics[width=\textwidth]{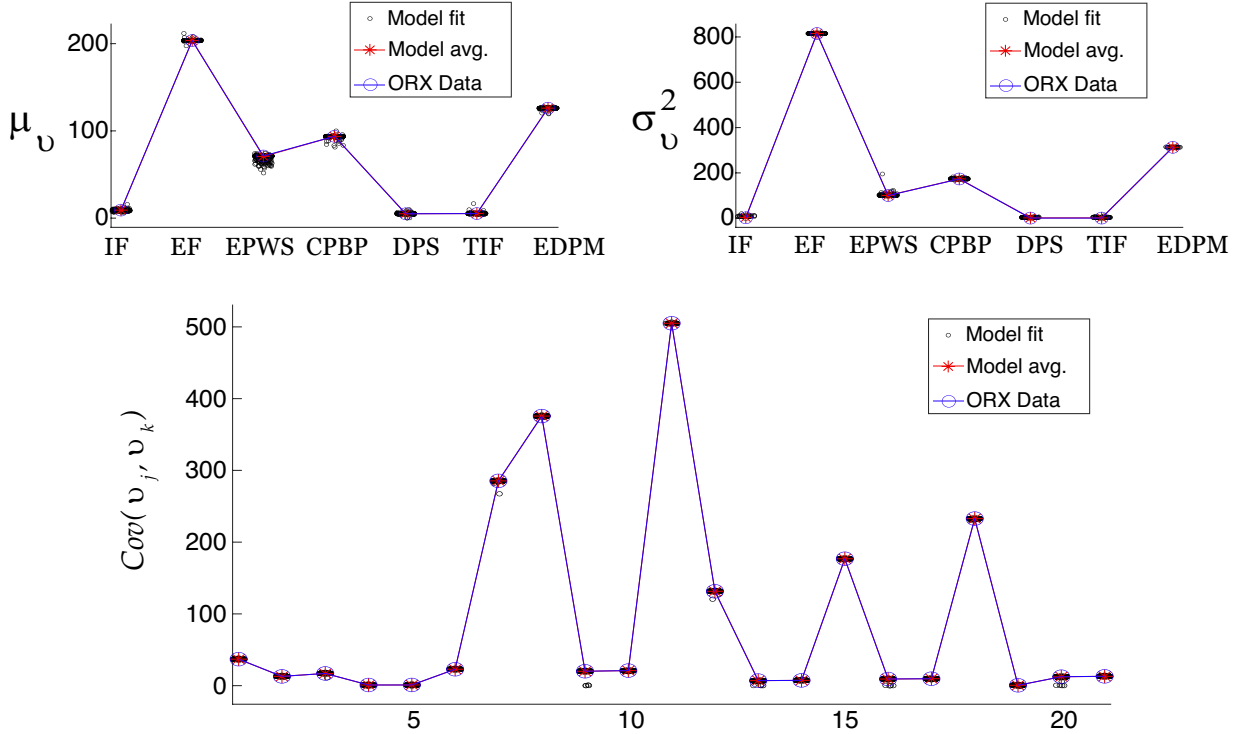}
    \caption{\label{fig:fit} \textbf{Fitting model to ORX data}. 
The 3 yearly frequency statistics we fit: mean $\mu_{\nu}$, variance $\sigma^2_{\nu}$, and covariance $Cov(\nu_j,\nu_k)$. 
Each of the black dots show the model fit to ORX data at a random starting point for the optimization routine (see \textbf{Materials and methods}), the 
red star is the average of all 1000 fits. In all instances, the model fits the ORX data (blue circles) very well, but recall the system is 
under-constrained with more parameters (42) than constraints (35).  The black circles are jittered horizontally for visual purposes.
}
\end{figure}

The model fits are very good (Fig \ref{fig:fit}) despite starting at 1000 random initial conditions using Latin hypercube sampling, see \textbf{Materials and methods} sub-section `Procedure for fitting model to ORX data' for details, 
and Fig. \ref{fig:Details} for parameter values. 
From Fig \ref{fig:fit}, we see that the solution to the system is not very sensitive to the random initial parameterization. 
This may not be surprising given that we have an under-constrained system of 35 equations and 42 unknowns (parameters), 
but note that parameters have constraints: $a_j,\tau_j,\gamma_j >0$ and $c_{j,k}\in(-1,1)$.  With more granular data, which we do not have access too, there would 
be more constraints that could certainly effect 
the quality of the model fits. Whether the solution is unique or not is unknown and beyond the scope of this current study.

\subsubsection{Effects of time window on covariance of loss distributions}

After finding parameters that capture the frequency distribution well, we can very quickly assess how the 
covariance $Cov(Q_j,Q_k)$ of the actual loss distributions varies with a large range of time windows; 
this would usually require time-consuming Monte Carlo simulations but our analytic calculations circumvent and simplify this.  
We use the average model fit (red stars in Fig \ref{fig:fit}) as the 42 parameters because there is little deviation among the 1000 model fits, the solution appears quite stable. 

\begin{figure}[!htb]
\centering
    \includegraphics[width=\textwidth]{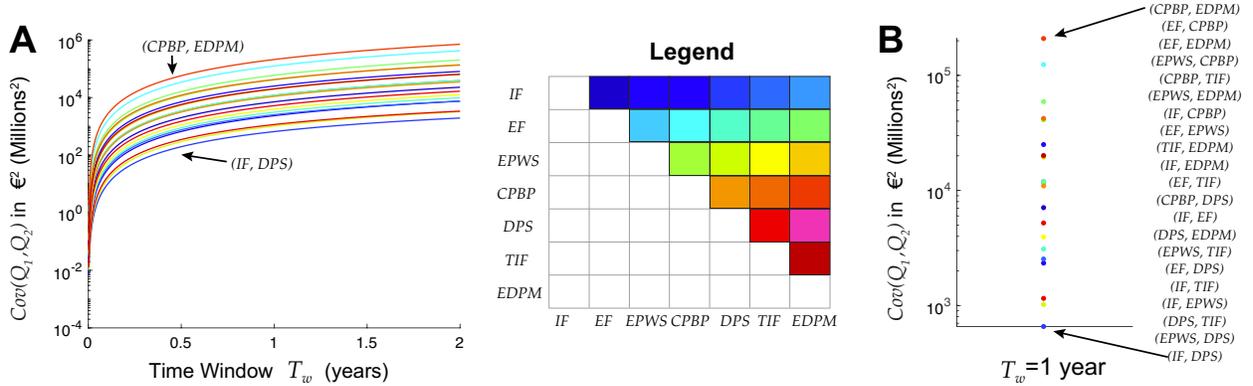}
    \caption{\label{fig:corx_varyT} \textbf{Covariance of loss distributions fit to ORX data}. 
\textbf{A}) The resulting $Cov(Q_j,Q_k)$ for all 21 pairs as a function of time window $T_w$ (log scale on vertical axes).  The 
statistics can vary a few orders of magnitude or more for these times. The legend is shown on the right. 
\textbf{B}) Shows the order of smallest and largest covariance pairs, setting $T_w=1$\,year for exposition purposes.  The ordering 
does not change as $T_w$ varies. 
}
\end{figure}

\begin{figure}[!htb]
\centering
    \includegraphics[width=.95\textwidth]{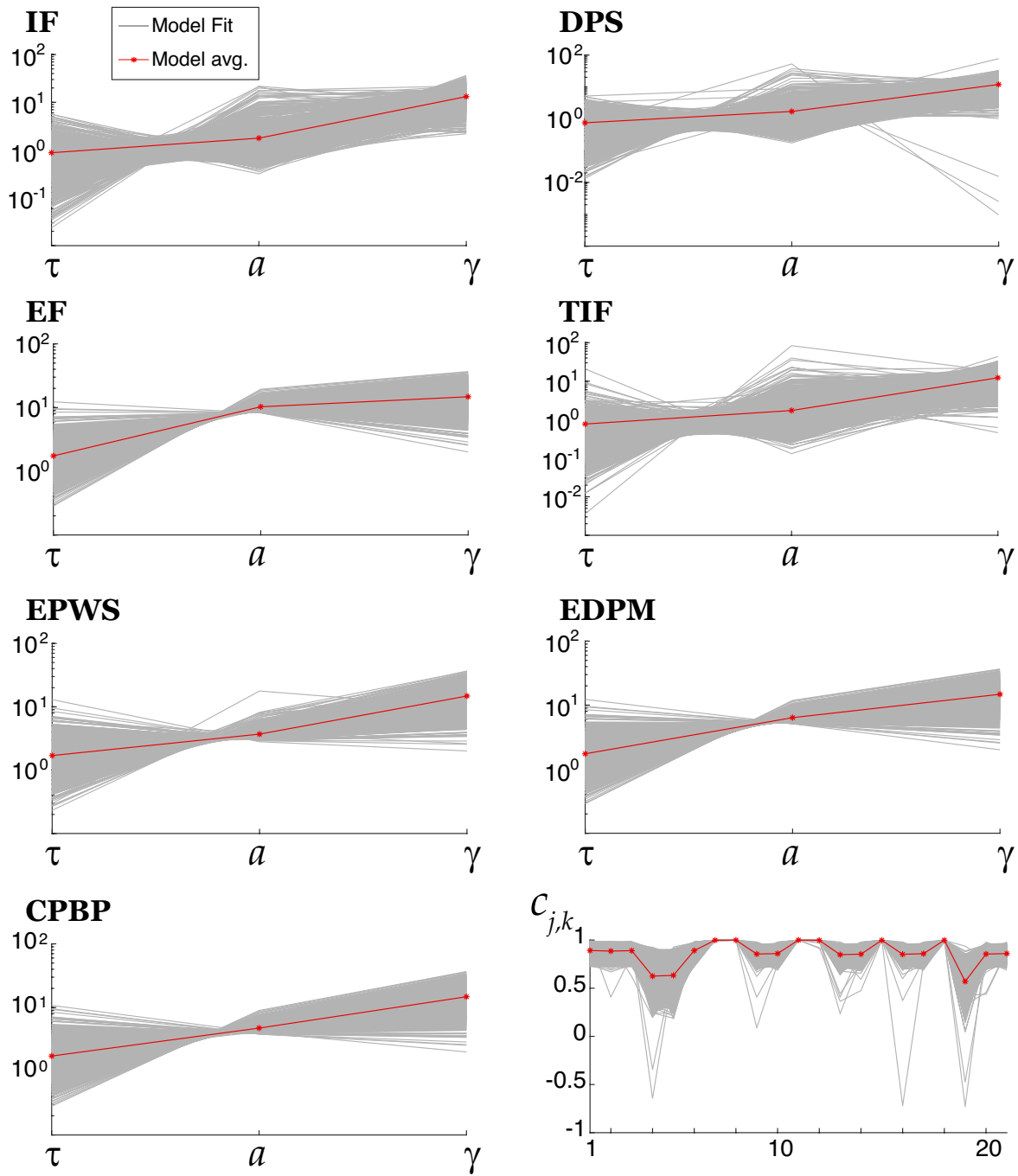}
    \caption{\label{fig:Details} \textbf{Details of 42 parameter values fit to ORX data.}. 
Showing all 1000 model fits (gray lines) with lines connecting a particular random parameterization, and the model averages (red stars).  The parameters 
$(\tau,a,\gamma)$ are segmented by the 7 risk categories, all 21 input correlations $(c_{j,k})$ are in lower right panel.
}
\end{figure}

The results are summarized in Figure \ref{fig:corx_varyT}.  Here we find that the largest covariance of $Q(t)$ is (CPBP, EDPM), and the smallest is with 
(IF,DPS).  We note that the univariate statistics for both the frequency and average severity per event (Table \ref{tbl:riskAvgs}) are not at all indicative of which 
pairs would have the highest covariance, and neither are the covariances of frequencies in Fig \ref{fig:covORX}A (i.e., the largest/smallest 
covariance in frequencies do not correspond to the largest/smallest covariance of losses in Fig \ref{fig:corx_varyT}).  This shows that 
the correlations of the aggregate loss distributions, specifically which pairs have the largest or smallest correlations, cannot intuited a priori and require proper modeling.

The results in Figure \ref{fig:corx_varyT} indicate that for all 21 covariances, care must be taken; systems 
should use a consistent time window, otherwise the covariances can vary by at least a few orders of magnitude when treating daily 
statistics as yearly statistics.

\section{Discussion}

We present, develop and apply a framework to mathematically account for how the correlation of loss distributions could vary when 
data are collected and/or processed on different time scales.  
Specifically, we use a dynamic stochastic differential equation model for the frequency distribution of Operational Risk loss events.  
This model is a generalization of commonly used homogeneous Poisson Process frequency model, capturing time-varying changes 
in the statistics of the frequency and equipped with temporal correlations that do not exist in homogeneous Poisson Process models.  
The frequency distribution model contains these complexities yet has only a few parameters and is amenable to mathematical calculations 
that result in relatively accurate formulas. This paper provides detailed mathematical calculations and thorough comparisons with 
Monte Carlo simulations with a large range of parameters. However, the method and overall framework we have outlined and implemented where the variance/covariance in arbitrary 
time windows is obtained from integrating the autocovariance/cross-covariance multiplied by $\Big(T_w - \vert t \vert \Big)$ (Eqs \eqref{eqn:autVar}, \eqref{eqn:cov_wind}), is general 
and does not require the specific inhomogeneous Poisson Process model for the frequency distribution we introduced. This method can be applied to 
any frequency distribution model as long as the auto/cross-covariance can be calculated.

Our application of methods relating the cumulative loss statistics to 
the statistics in smaller time windows is inspired by the successful usage in 
other disciplines (signal processing \citep{kay1993}, 
computational neuroscience \citep{lindner05,shea2008correlation,litwin2011balanced,litwin2012spatial,BarreiroLy_RecrCorr_17,BarreiroLy_jmns_18}), 
but the models we use here are specific to Operational Risk, thus yielding novel formulas and applications.  

The loss time series have temporal correlation within a risk category \textit{and} correlation across risk categories, so we use conventional Monte Carlo simulations with a time discretization. 
The time discretization is advantageous for obtaining the simulated auto/cross-covariance function needed in our theory (recall Figs \ref{fig:varyN}, \ref{fig:cov1}A,B).  
Whether it is possible to simulate just the event times without discretizing time (i.e., Gillespie algorithm) and rescale to equivalent pathwise deterministic Markov processes 
is not obvious to us; 
we believe this would require new methods and is beyond the scope of this study.

We demonstrate in principle how our loss distribution model could be fit to actual operational risk data. Since the data was provided by an 
exchange (ORX) with only industry-wide averages from many different institutions that vary in size, scope, objective, etc., the details of our application and 
results would likely be very different in practice. 
The ORX data was very coarse, lacking granular details of loss events that individual institutions might access to, including time of occurrence and magnitude.  
Nevertheless, the ORX data contained information about the average severity of losses per event, segmented by 7 common risk categories, as well as the
year-to-year frequency of events from 2014--2019.  Although not meant to be a definitive model to supplant the status quo, 
we use the model to estimate the covariances of aggregate loss distributions.  Moreover, our mathematical calculations enabled us to quickly assess how the 
covariances of the losses change with different time windows, showing that if institutions are not careful in their systems, e.g., 
they are are not using the same time windows to estimate covariances (especially with risk categories with inherently different time-scales) 
that feed into copula calculations, the results of their capital calculations could be significantly inaccurate. 

We close with some remarks about applying our method to actual data. Institutions use their proprietary internal data that has significantly more details of individual loss events, 
so they can certainly have more data constraints in their model fitting than what we demonstrated here. It is natural to have an over-constrained system in fitting our model 
to an institution's data. Individual institutions also do not necessarily use these specific seven risk categories provided by ORX (Table \ref{tbl:riskAbbrev}), they may have more (and different) risk categories that could each be modeled by a different frequency and severity model. In our specific demonstration, we did not normalize the magnitudes of the ORX data across risk categories; it may be desirable to model relative percent differences versus capturing vast differences across risk categories. Sensitivity analyses of model fits are a key component: with our model, the mathematical formulas used to estimate (co-)variances should be varied (perhaps with Monte Carlo simulations, and also setting the correlations to extreme values of $\pm 1$) to assess how capital calculations change. The merits of our model applied to a specific institution and their data are to be determined, but note that all model components in this paper, including figures and tables, are freely available on GitHub. 

\section{Materials and methods}

See GitHub page \url{https://github.com/chengly70/OperationalRisk/} for freely available code simulating the models in this paper, including comparisons 
to Monte Carlo simulations, implementation of analytic formulas, processing of freely available ORX data, model fitting to ORX data, etc.

For the severity distribution we consider several commonly used parametric distributions $f_S(x)$ \citep{chern08book}:
\begin{align} 
    f_S(x) &= \frac{1}{\Gamma(\alpha)\beta}\left( \frac{x}{\beta} \right)^{\alpha-1} \exp\Big(-x/\beta \Big), \text{  for }x>0; \alpha,\beta>0, 
    \text{  Gamma} \\
    f_S(x) &= \frac{1}{x \sqrt{2\pi}\sigma}\exp\Big(-\frac{(\log(x)-\mu)^2}{2\sigma^2} \Big), \text{  for }x>0; \mu\in\mathbb{R}, \sigma>0, 
    \text{  Lognormal} \\
    f_S(x) &= \frac{1}{\sigma}\left(1+k\frac{x}{\sigma}\right)^{-1-1/k}, \text{ for }x> 0, \text{  GPD}\\ 
    f_S(x) &= \frac{b}{a}\left(\frac{x}{a}\right)^{b-1}\exp\Big(-\left(\frac{x}{a}\right)^b\Big), \text{ for }x>0;b>0;a>0, \text{ Weibull}\\
    f_S(x) &= \frac{\frac{kc}{\alpha}\Big(\frac{x}{\alpha}\Big)^{c-1}}{\Big(1+\Big(\frac{x}{\alpha}\Big)^{c}\Big)^{k+1}}, \text{ for }x>0;\alpha>0;c>0;k>0, \text{ Burr} 
\end{align}

The statistics for these common distributions are shown in Table \ref{tbl:stats}.  

\begin{table}[hbt]
\caption{\label{tbl:stats} Severity distribution family and statistics.  Here $\Gamma(x)=\int_0^\infty z^{x-1}e^{-z}\,dz $ and 
$B(x,y)=\Gamma(x)\Gamma(y)/\Gamma(x+y)$.  
To insure the mean and variances are finite, in \textbf{GPD} we have $0\leq k <\frac{1}{2}$, in \textbf{Burr} we have $k>\frac{2}{c}$.}
\noindent\adjustbox{max width=\textwidth}{ 
\begin{tabular}{l|l|l|l}
\multicolumn{1}{c|}{\textbf{Distribution}} & \multicolumn{1}{c|}{\textbf{Mean $\mathbb{E}[S]=\mu_S$}} & \multicolumn{1}{c|}{\textbf{Variance} $\mathbb{E}[S^2]-\mathbb{E}[S]^2=\sigma^2_S$} & \multicolumn{1}{c}{\textbf{Allowable Parameters}} \\ \hline\hline
\textbf{Gamma}                             & $\alpha \beta$                                     & $\alpha\beta^2$  &    $\alpha,\beta>0$                 \\ \hline
\textbf{Lognormal}                         &   $\exp\left(\mu+\sigma^2/2 \right)$                   &    $\left(e^{\sigma^2}-1\right)e^{2\mu+\sigma^2}$ &  $\sigma>0$ \\ \hline
\textbf{GPD}                               & $\frac{\sigma}{(1-k)}$                                 &  $\frac{\sigma^2}{(1-k)^2(1-2k)}$   
& $\sigma>0$ and $0\leq k < \frac{1}{2}$ 
\\ \hline
\textbf{Weibull}                           & $a\Gamma\Big(1+\frac{1}{b}\Big)$                             & $a^2\Gamma\left(1+\frac{2}{b}\right)-a^2\Big(\Gamma\left(1+\frac{1}{b} \right)\Big)^2$              &   $a,b>0$                     \\ \hline
\textbf{Burr}                              & $k\alpha B\left(k-\frac{1}{c}, 1+\frac{1}{c}\right)$           & $k\alpha^2B\left(k-\frac{2}{c}, 1+\frac{2}{c}\right)-(k\alpha)^2B\left(k-\frac{1}{c}, 1+\frac{1}{c}\right)^2$     &   $\alpha,c,k>0$ and $k>\frac{2}{c}$                              
\end{tabular}
}
\end{table}

\subsection{Calculations for a single inhomogeneous Poisson Process frequency model}

Recall $R(t) = S * I(t)$, with $P(I(t)=1)=\nu(t)\Delta t$ where the time points $t$ are spaced a part by $\Delta t$, 
and where $\nu(t)$ is the probability per unit time of a loss event occurring.  The model for $\nu(t)$ is:
\begin{equation}
 \tau \frac{d}{dt}\nu(t)=-\nu(t) + \tau a \sum_{k} \delta(t-t_k)   
\end{equation}
where $t_k$ are random points drawn from a homogeneous Poisson Process with rate $\gamma$, 
$a$ is the jump size of $\nu(t)$ at times $t_k$, and $\tau$ is the time-scale that determines how fast $\nu(t)$ decays to 0 
in the absence of random jumps at $t_k$.

The statistics of $\nu(t)$ are calculated by a stochastic integral formulation of the equation: 
$$\nu(t) = a \int_{0}^{\infty} D(t-t')e^ \frac{-t'}{\tau}\,dt' $$
where $D(t) = \displaystyle\sum_{k} \delta(t-t_k) $, and $\mathbb{E}[D(t)]=\gamma$.  
Taking the expected value of this equation and recognizing the only source of randomness on the rhs is $D$, we get:
$$\mathbb{E}\left[ {\nu(t)} \right]=a\int_{0}^{\infty} \mathbb{E}\left[D(t-t')\right]e^ \frac{-t'}{\tau}\,dt' = a\gamma \tau. $$ 
Thus we have: 
\begin{equation} \label{eqn:mnRinhomog}
    \mathbb{E}\left[R\right] = \mathbb{E}_\nu \Big[ \mathbb{E}[R\vert \nu] \Big]= 
\mathbb{E}_{\nu(t)}[ \mu_S \nu(t) \Delta t ] = \mu_S\Delta t \mathbb{E}_{\nu(t)}[\nu(t)]= \mu_S(a\gamma \tau \Delta t). 
\end{equation}
For the variance we need the $2^{nd}$ moment of $\nu(t)$.
\begin{equation}\label{eqn:nut2int}
 \nu(t)^2=a^2\int_{0}^{\infty}\int_{0}^{\infty}D(t-u)D(t-v)e^ \frac{-u}{\tau}e^ \frac{-v}{\tau} \,du\,dv.
\end{equation}
Using this property of homogeneous Poisson Processes: 
$$\mathbb{E}\left[D(t-u)D(t-v)\right]=(a\gamma )^2+\gamma \delta(u-v),$$ 
we get the second moment of $\nu(t)$ is:
\begin{equation}\label{eqn:secmnu}
    \mathbb{E}\left[\nu(t)^2\right]= (a\gamma\tau)^2+a^2\gamma\int_{0}^{\infty}e^\frac{-2v}{\tau}\,dv=
(a\gamma\tau)^2+\frac{a^2\gamma\tau }{2}.    
\end{equation}
The variance of $\nu(t)$ is:
\begin{equation}\label{eqn:nuTvar}
    Var(\nu(t)) = \frac{a^2\gamma\tau}{2}
\end{equation}
The autocovariance $\mathbb{E}[\nu_{t'}\nu_{t'+t}]-\mathbb{E}[\nu(t)]^2$ is similarly calculated by replacing $D(t-u)$ with 
$D(t'+t-u)$ in Eq \eqref{eqn:nut2int}:
\begin{equation}\label{eqn:nut2}
    A_{\nu(t)}(t) = \frac{a^2\gamma\tau}{2} e^{-\vert t \vert / \tau}
\end{equation}

Since the second moment of $R$ is: 
$$\mathbb{E}_\nu\Big[ \mathbb{E}[ R^2 \vert \nu(t) ]\Big] = \mu_{S^2}\Delta t\mathbb{E}_{\nu(t)}[ \nu(t)] = \mu_{S^2}(a\tau\gamma\Delta t), $$
the variance of $R$ is:
\begin{equation}\label{eqn:inhomVarR}
    \sigma^2_R = \mu_{S^2}(a  \tau \gamma \Delta t) - \left(\mu_S a \tau \gamma \Delta t\right)^2
\end{equation}

For cumulative losses in larger time windows, the statistics are calculated similarly, with: 
\begin{equation}\label{eqn:mnSever_inhom}
    \mu_{Q} = \sum \mathbb{E}[R] = n\Delta t \Big( a\gamma\tau \mu_S \Big)= T_w a\gamma\tau \mu_S
\end{equation}

To obtain the autocovariance of $R$, we leverage that fact that the support (where $R>0$ in time) is the same as the realization 
of $\nu(t)$, so the temporal structure of $R$ and $\nu(t)$ are identical, with the only difference being the magnitudes.  So the autocovariance of $R$ must also be of similar form to $A_{\nu(t)}$ 
except we have $A_{R}(t=0)=\sigma^2_R$:
\begin{equation}\label{eqn:autcR}
    A_R(t) = \sigma^2_R \frac{ e^{-\vert t \vert / \tau} }{\Delta t}.
\end{equation}

With the calculation of $A_R(t)$ in hand, we use 
$$ \sigma^2_{Q} = \int_{-T_w}^{T_w} A(t)\Big( T_w - \vert t \vert \Big)\,dt $$
(repeat of equation \eqref{eqn:autVar}) to get an \underline{\textit{approximation}} to the 
variance of cumulative losses in a time window $T_w$:
\begin{equation}\label{eqn:inhomVarRsc}
    \sigma^2_{Q}  = \int_{-T_w}^{T_w} A_R(t)\Big( T_w - \vert t \vert \Big)\,dt = 
    2\frac{\sigma_R^2}{\Delta t} \tau \Big( T + \tau(e^{-T/\tau} -1 ) \Big). 
\end{equation}
This approximation assumes $A(t)$ suffices to capture $\sigma^2_{Q}$. 

\subsection{Calculations for two loss time series}

To generalize the theory and calculations from the prior section, we consider the cross-covaraince function of 
$CC_{\nu}(t):=\mathbb{E}[\nu_1(t'+t)\nu_2(t')]-\mathbb{E}[\nu_1]\mathbb{E}[\nu_2]$; 
note that $CC_{\nu}(t)\neq CC_{\nu}(-t)$, 
unlike with 
the autocovariance function where $A(t)=A(-t)$ for both $R$ and $\nu(t)$.  
The analogous equation to Eq. \eqref{eqn:nut2int} is:
\begin{equation}\label{eqn:secnu12}
    \nu_1(t')\nu_2(t'+t) = a_1 a_2\int_{0}^{\infty}\int_{0}^{\infty}D_1(t'+t-u)D_2(t'-v)e^ \frac{-u}{\tau_1}e^ \frac{-v}{\tau_2} \,du\,dv.
\end{equation}
where $D_j(t) = \displaystyle\sum_{k_j} \delta(t-t_{k_j}) $.  Since 
$$\mathbb{E}[D_1(t'+t-u)D_2(t'-v)] - \gamma_1 \gamma_2 = c\bar{\gamma} \delta(t-u+v)  $$ 
by construction, 
we take the expected value of Eq. \eqref{eqn:secnu12}, using similar calculations as before (cf. Eq. \eqref{eqn:secmnu}) to get:
\begin{equation}\label{eqn:crssnu}
    CC_{\nu}(t)= c \bar{\gamma} a_1 a_2 \frac{\tau_1 \tau_2}{\tau_1+\tau_2} \left\{ \begin{array}{cc} 
                e^{-t/\tau_1}, & \text{ if   } t\geq 0 \\
                e^{-\vert t\vert/\tau_2}, & \text{ if   } t<0
                \end{array} \right. .
\end{equation}

We apply the same arguments as before when deriving $A_R(t)$ (Eq. \eqref{eqn:autcR}): since $R(t)$ and $D_j$ 
have the same temporal support, the cross-covariance functions must be of a similar form.  
We first derive the 
point-wise covariance for the losses in small windows $\Delta t$: $Cov(R_1,R_2)$ and set this to 
$CC_R(t=0)$ to get:
\begin{equation}\label{eqn:CCr}
CC_R(t) =  \frac{Cov(R_1,R_2)}{\Delta t} \left\{ \begin{array}{cc} 
                e^{-t/\tau_1}, & \text{ if   } t\geq 0 \\
                e^{-\vert t\vert/\tau_2}, & \text{ if   } t<0
                \end{array} \right. 
\end{equation}

To derive $Cov(R_1,R_2)$, we employ similar methods for $\sigma^2_R$ (Eq. \ref{eqn:inhomVarR}):
\begin{eqnarray}
    \mathbb{E}[R_1(t)R_2(t)] &=& \mathbb{E}_\nu\Big[ \mathbb{E}[ R_1 R_2 \vert \nu ]\Big] \nonumber \\  
                            &=& \mu_{S_1}\mu_{S_2} \mathbb{E}_\nu[\nu_1\nu_2] (\Delta t)^2 \nonumber \\
                            &=& \mu_{S_1}\mu_{S_2} \Big(c\bar{\gamma}a_1 a_2\frac{\tau_1\tau_2}{\tau_1+\tau_2}(\Delta t)^2+(a_1\tau_1\gamma_1 \Delta t)(a_2\tau_2\gamma_2 \Delta t) \Big),
\end{eqnarray}
to get:

\begin{equation}\label{eqn:covR}
    Cov(R_1,R_2) =  \mu_{S_1}\mu_{S_2} c\bar{\gamma} a_1 a_2\frac{\tau_1\tau_2}{\tau_1+\tau_2} (\Delta t)^2.
\end{equation}
Using the same method as before to relate the autocovariance of $R$ to the variance of cumulative losses in larger time windows $T_w$, 
we relate the cross-covariance function to the covariance of cumulative losses in $T_w$:
\begin{eqnarray}
    Cov(Q_1,Q_2) &=& \int_{-T_w}^{T_w} CC_R(t)\Big( T_w-\vert t \vert \Big)\,dt \nonumber \\
                                &=& Cov(R_1,R_2) \frac{\tau_1\Big(T_w+\tau_1(e^{-T_w/\tau_1}-1)\Big) +\tau_2\Big(T_w+\tau_2(e^{-T_w/\tau_2}-1)\Big)}{\Delta t} \nonumber
\end{eqnarray}
\begin{equation}  \label{eqn:covTw1} 
    Cov(Q_1,Q_2)=c\bar{\gamma} \mu_{S_1}\mu_{S_2}  a_1 a_2\frac{\tau_1\tau_2}{\tau_1+\tau_2} \Big[ \tau_1\Big(T_w+\tau_1(e^{-T_w/\tau_1}-1)\Big) + \tau_2\Big(T_w+\tau_2(e^{-T_w/\tau_2}-1)\Big) \Big]\Delta t
\end{equation}

\subsection{Details for Obtaining ORX Data}
 
 The total frequency and severity of operational risk losses from the consortium of all banks, segmented 
 by risk categories, is directly reported in the ORX Annual Banking Lost Report \citep{orx_urlData1} 
 (see Tables \ref{tbl:orxfreq}--\ref{tbl:orxsev}).  
 There are however two important aspects of the data that we must address so that it is in a usable 
 form (i.e., yearly data for Banking losses only): i) the data is cumulative over 6 years (2014--2019), 
 ii) the data combines all business types (Banking, Trading and Investments, Corporate Items, Other) 
 while our application concerns only Banking losses. 
 
 \begin{table}[h!]
\caption{ \label{tbl:orxfreq} Extracted data of frequency of events with risk category proportions, 
from page 11 of \citet{orx_urlData1}. We calculated the row `\% of Total from Bank Loss' as a proportion of the total, 
the last row is from page 10 of \citet{orx_urlData1}.}
\noindent\adjustbox{max width=\textwidth}{ 
\begin{tabular}{l|ccccccc|c}
& IF & EF & EPWS & CPBP & DPS & TIF & EDPM & Totals                  \\ \thickhline
Retail Banking                                    & 4561                                                     & 100361                                                     & 35026                                                                                                  & 39991                                                                                                & 2509                                                                                 & 2116                                                                                           & 51352                                                                                                   & 235916                                       \\
Private   Banking                                 & 140                                                      & 1671                                                       & 1794                                                                                                   & 3268                                                                                                 & 23                                                                                   & 110                                                                                            & 3947                                                                                                    & 10953                                        \\
Commercial   Banking                              & 327                                                      & 8949                                                       & 2033                                                                                                   & 7891                                                                                                 & 206                                                                                  & 631                                                                                            & 13462                                                                                                   & 33499                                        \\ \hline
Totals                                            & 5028                                                     & 110981                                                     & 38853                                                                                                  & 51150                                                                                                & 2738                                                                                 & 2857                                                                                           & 68761                                                                                                   & 280368                                       \\ \hline
\% of Total from Bank Loss                           & 2\%                                                      & 40\%                                                       & 14\%                                                                                                   & 18\%                                                                                                 & 1\%                                                                                  & 1\%                                                                                            & 25\%                                                                                                    &                     \\
\% Bank Losses to Gross & 73\%                             & \multicolumn{1}{l}{}               & \multicolumn{1}{l}{}                                                           & \multicolumn{1}{l}{}                                                         & \multicolumn{1}{l}{}                                         & \multicolumn{1}{l}{}                                                   & \multicolumn{1}{l}{}                                                            & \multicolumn{1}{l}{}
\end{tabular}
}
\end{table}

\begin{table}[h!]
\caption{ \label{tbl:orxsev} Extracted data of severity of losses (\euro{}-Million) 
with risk category proportions, from page 13 of \citet{orx_urlData1}. We calculated the row `\% of Total from Bank Loss' as a proportion of the total, 
the last row is from page 12 of \cite{orx_urlData1}.}
\noindent\adjustbox{max width=\textwidth}{
\begin{tabular}{l|ccccccc|c}
& IF & EF & EPWS & CPBP & DPS & TIF & EDPM & Totals                  \\ \thickhline
Retail Banking                                    & 950                                                      & 7045.2                                                   & 3394.7                                                                                            & 26504.3                                                                                         & 389.7                                                                            & 844.4                                                                                      & 13655.6                                                                                             & 52783.9                                      \\
Private Banking                                   & 161.1                                                    & 304.4                                                    & 233.8                                                                                             & 2456.2                                                                                          & 2.1                                                                              & 27.6                                                                                       & 886.6                                                                                               & 4071.8                                       \\
Commercial  Banking                               & 158.4                                                    & 4514                                                     & 240                                                                                               & 12962                                                                                           & 30.3                                                                             & 678.4                                                                                      & 5425.1                                                                                              & 24008.2                                      \\ \hline
Totals                                            & 1269.5                                                   & 11863.6                                                  & 3868.5                                                                                            & 41922.5                                                                                         & 422.1                                                                            & 1550.4                                                                                     & 19967.3                                                                                             & 80863.9                                      \\ \hline
\% of Total from Bank Loss                             & 2\%                                                      & 15\%                                                     & 5\%                                                                                               & 52\%                                                                                            & 1\%                                                                              & 2\%                                                                                        & 25\%                                                                                                &                      \\
 
\% Bank Losses to Gross & 56\%                             &                                                          &                                                                                                   &                                                                                                 &                                                                                  &                                                                                            &                                                                                                     &                                             
\end{tabular}}
\end{table}

 Starting with Tables \ref{tbl:orxfreq}--\ref{tbl:orxsev}, we approximate the yearly frequency and severity 
 by taking a fraction (73\% for frequency, 56\% for severity) 
 of the yearly amounts reported on page 6 of \citet{orx_urlData1} (see Table \ref{tbl:orxyear}).  
 Note that the percentages (73\%,56\%) are the reported fractions from Bank losses alone; see 
 pages 10 and 12, respectively, in \citet{orx_urlData1}. With an approximation for each of the 6 years, 
 we then use the computed percentages (row 5 of Tables \ref{tbl:orxfreq}--\ref{tbl:orxsev}) to distribute 
 the loss data by risk category.  The final step is to divide the loss data by the number of 
 contributing bank institutions (see bottom row of Table \ref{tbl:orxyear}), 
 to get yearly data of realized losses from the average institution.

 \begin{table}[!htb]
 \centering
 \caption{ \label{tbl:orxyear} Extracted frequency and severity data by year (page 6 of \citet{orx_urlData1}) 
 that includes all business types: Banking, Trading and Investments, Corporate Items, Others.  
 The total number of institutions reporting by year (bottom row) is from page 4 of \citet{orx_urlData1}). }
\noindent\adjustbox{max width=\textwidth}{
\begin{tabular}{l|cccccc}
Year                                            & 2014  & 2015  & 2016  & 2017  & 2018  & 2019 \\ \hline
Total \# Events                                 & 65766 & 69519 & 65797 & 63491 & 59642 & 59437                     \\
Total Losses (\euro{}--Billion) & 37.6  & 25.1  & 28.1  & 20    & 17.6  & 15.8                      \\
Total \# Institutions                           & 80    & 85    & 92    & 96    & 97    & 100                      
\end{tabular}
}
\end{table}

\subsection{Procedure for fitting model to ORX data}

Recall the system of 35 equations is:
\begin{eqnarray}
	a_j \tau_j \gamma_j T_w &=& \mu_{\nu_j,ORX}   \text{   ; for }j=1,\dots,7  \nonumber \\
	a_j^2\gamma_j \tau_j^2 \Big( T_w + \tau_j(e^{-T_w/\tau_j} -1 ) \Big) &=& \sigma^2_{\nu_j,ORX}    \text{   ; for }j=1,\dots,7  	\nonumber 
\end{eqnarray}
\begin{eqnarray}	
	c_{j,k}\bar{\gamma}  a_j a_k\frac{\tau_j\tau_k}{\tau_j+\tau_k} \Big[ \tau_j\Big(T_w+\tau_j(e^{-T_w/\tau_j}-1)\Big) +\tau_k\Big(T_w+\tau_k(e^{-T_w/\tau_k}-1)\Big) \Big] =   \nonumber \\
	 Cov(\nu_{j,ORX},\nu_{k,ORX}) \text{   ; for }j\neq k \nonumber
\end{eqnarray}
the 42 unknowns are $(a_j,\tau_j,\gamma_j, c_{j,k})=:\vec{x}$. The left-hand side are statistics from the model $X_{mod}$ and the right-hand side are the 
statistics from ORX data. 

We use MATLAB's {\tt fmincon} optimization routine to find parameters that minimize the following objective function, i.e., the $l_1$ norm of data and model:
\begin{equation}
	\min_{\vec{x}} \sum_{j=1}^{7} \Big( \left\vert \mu_{\nu_j,mod}-  \mu_{\nu_j,ORX} \right\vert + \left\vert \sigma^2_{\nu_j,mod} - \sigma^2_{\nu_j,ORX} \right\vert \Big)
	+ \sum_{j<k} \left\vert  Cov(\nu_{j,mod},\nu_{k,mod})- Cov(\nu_{j,ORX},\nu_{k,ORX}) \right\vert
\end{equation}

We do not know if there exists a unique solution to this problem, so we use {\tt fmincon} with 1000 random initial parameterizations $\vec{x}_0$ using Latin hypercube sampling with:
\begin{eqnarray} 
	a_j &\sim & Unif(0,15) \nonumber \\
	\tau_j &\sim & Unif(0,2) \nonumber \\
	\gamma_j &\sim & Unif(0,40) \nonumber \\
	c_{j,k} &\sim & Unif(-1,1) \nonumber 
\end{eqnarray}
These particular distributions were arbitrarily chosen, except for the input correlation because $-1<c_{j,k}<1$ and the constraint that $a,\tau,\gamma>0$. 
The upper bounds for $(a,\tau,\gamma)$ are within the range of average number of events per year across the 7 risk categories (see Table \ref{tbl:riskAvgs}). 

The code to implement all of this is freely available at 

\noindent
\url{https://github.com/chengly70/OperationalRisk/} in function {\tt MORE\_fitFreq\_Parms.m}.

\section{Appendix: Homogenous Poisson Process} \label{sec:homog}


We first consider a simple time series where the frequency distribution is given by a Poisson Process with 
rate $\nu_r$ events/time, and an independent severity distribution $f_S$.  
Note that despite the simplicity of this model, it is still commonly used as a default model, especially for loss categories 
without a lot of data, because it is relatively transparent \citep{chern08book}. 
We present the calculation of second order statistics of the whole time series for a given loss category. 
For each risk type and a given period, operational losses are represented by a time series $R_j$, where $j$ indexes a time interval of size $\Delta t$ (assumed 
to be small enough so that at most one operational loss event can occur); this particular aggregate loss distribution is commonly referred to as a marked Poisson Process \citep{pplect17}. 

The probability of a loss event at a given time bin $(t,t+\Delta t)$ is $\nu_r \Delta t$ 
and each time bin is independent. To derive desired first and second order statistics, we use the PDF of $R$, where 
\begin{eqnarray}
P(I=1) = \nu_r\Delta t, & P(I=0)=1-\nu_r\Delta t .
\end{eqnarray}

The mean is: 
\begin{eqnarray}
	\mu_R &=&  \mathbb{E} [  I S ] = \mathbb{E}[I]\mathbb{E}[S] \nonumber \\
    &=& 0*P(I=0) \mathbb{E}[S] + 1*P(I=1) \mathbb{E}[S]  = 1*\nu_r\Delta t \mu_S\nonumber \\
    \Rightarrow \mu_R &=& (\nu_r \Delta t)\mu_S \label{eqn:meanR_homo}
\end{eqnarray}
where $\mu_S$ is the mean of the chosen severity distribution.  The second moment of $R$ is: 
$$ \mathbb{E}[R^2] = 0^2 *(1-\nu_r \Delta t)+ 1^2*(\nu_r\Delta t) \mathbb{E}[S^2] = (\nu_r \Delta t)\mu_{S^2} $$
so the variance is:
\begin{equation}\label{eqn:varR_homo}
        \sigma^2_R = (\nu_r \Delta t)\mu_{S^2}-(\nu_r \Delta t\mu_S)^2
\end{equation}

\subsubsection{Cumulative Loss Statistics in Arbitrary Time Windows}\label{sec:homoTwin}

For the purposes of aggregating capital over different time horizons (i.e., yearly capital assessment is 
a regulatory requirement \citep{bis_2006,chern08book}), we consider cumulative losses over different 
time windows to understand how this practice of using different time windows might effect the statistics.  Development of methods to capture cumulative losses in arbitrary time windows may 
also help institutions handle certain operational loss categories that occur infrequently or 
where data quality is bad and thus data is unreliable.  Recall that $R$ is the loss in small time bin $\Delta t$, 
the cumulative losses in a time window of length $T_{w}$ that contains $n:=T_w/\Delta t$ time bins is:
\begin{equation} \label{eqn:cumR_homo}
    Q_l = \sum_{j=1}^{n} R_{j+(l-1)n} 
\end{equation}
where the subscript $l$ denotes the $l^\text{th}$ window of length $T_w$.  
The mean is:
\begin{equation}\label{eqn:mnSever_pp}
    \mu_{Q} = \sum \mathbb{E}[R] = n\Delta t \Big( \nu_r \mu_S \Big)= T_w \nu_r \mu_S
\end{equation}
With a homogeneous Poisson Process model, each of the $R$'s are independent, making the calculation of the variance 
simpler because the covariance terms are 0:
\begin{eqnarray}
    \sigma^2_{Q} = \sum Var(R_j)+2\sum_{j<k} Cov(R_j,R_k) &=& n \left[(\nu_r \Delta t)\mu_{S^2}-(\nu_r \Delta t\mu_S)^2  \right] \nonumber \\
    & = & T_w \nu_r \Big( \mu_{S^2} - \nu_r\Delta t (\mu_S)^2 \Big)
    \label{eqn:varSever_pp}
\end{eqnarray}

\begin{figure}[!htb]
\centering
\includegraphics[width=.95\columnwidth]{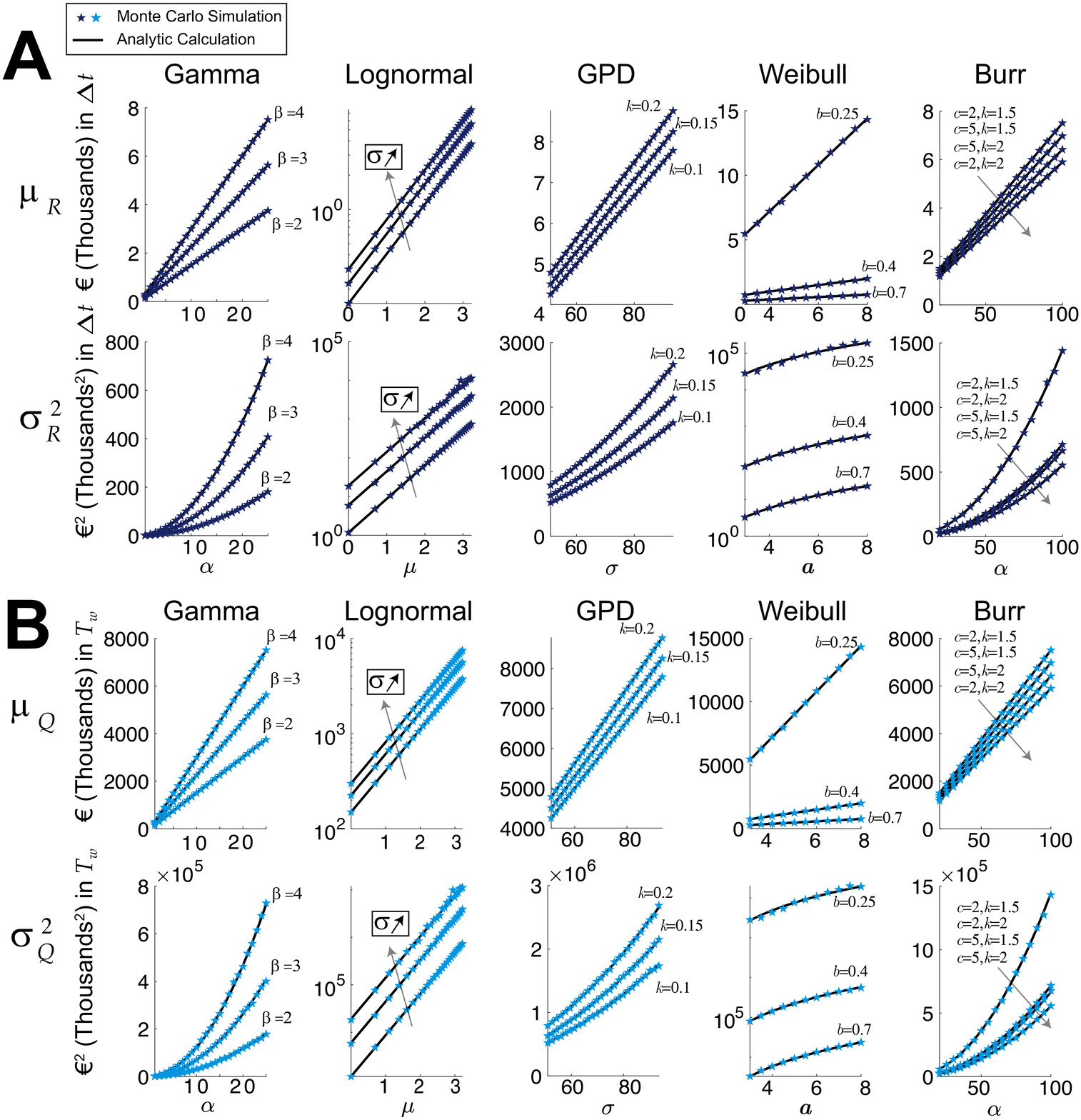}
\caption{ \label{fig:homoCompare} \textbf{Accurate analytic formulas for homogeneous PP frequency model.} 
Comparing the mean and var of $R$ and $Q$ following format of Figure \ref{fig:hetCompare} but 
with a homogeneous Poisson Process 
frequency distribution ($\nu_r=75$ events/year) for 
common severity distributions (see Table \ref{tbl:stats}).  Note that Table \ref{tbl:riskAvgs} shows that 
this average frequency is typical in industry-wide reporting.  
\textbf{A}) Mean $\mu_R$ and variance $\sigma^2_R$ of aggregate loss distribution 
in small windows $\Delta t=0.001$ in time units of years ($\approx 0.365$ day), 
with Monte Carlo simulations in stars (Eq. \eqref{eqn:R_homo}) and analytic formulas in solid curves 
(Eq. \eqref{eqn:meanR_homo} and \eqref{eqn:varR_homo}).  
\textbf{B})  Mean $\mu_Q$ and variance $\sigma^2_Q$ of cumulative losses 
in large time windows $T_w=1$ year, 
with Monte Carlo simulations in stars (Eq. \eqref{eqn:cumR_homo}) and 
calculations with solid curves (Eq. \eqref{eqn:mnSever_pp} and \eqref{eqn:varSever_pp}).  
}
\end{figure}

\subsubsection{Relationship between autocovariance and $Q$}

The autocovariance function is common tool to quantify the temporal dynamics of a time series; here we define it as:
\begin{equation}\label{eqn:autoDefn}
    A(t) = \mathbb{E}_{t'}\left[ R_{t'+t}R_{t'} \right] - \mathbb{E}_{t'}\left[R_{t'} \right]^2
\end{equation}

$A(t)$ specifies how correlated (or degree of co-variability) 
two points separated by time $t$ are. 
Note that $A(t=0)=\sigma^2_T$ is the (point-wise) variance calculated from samples at each $\Delta t$. 
The autocovariance has a nice relationship with the cumulative cum in a window of length $T_w$ whenever the time series satisfies stationarity \citep{kay1993}: 
\begin{equation}
    \sigma^2_{Q} = \int_{-T_w}^{T_w} A(t)\Big( T_w - \vert t \vert \Big)\,dt
\end{equation}

Indeed, we can use this equation to derive equation \eqref{eqn:varSever_pp}; note that the Autocovariance function of 
$R$ with a Poisson Process frequency distribution and independent severity distribution is simply:
\begin{equation}\label{eqn:autPPmod}
    A(t) = \sigma^2_R \delta(t) = \sigma^2_R \frac{1}{\Delta t}.
\end{equation}

Substituting this into equation \eqref{eqn:autVar}, we get:

$$\sigma^2_{Q} = \sigma^2_{R} \Big(T_w-\vert t \vert \Big)\Big\vert_{t=0} = \sigma^2_{R} T_w = 
\Big( \nu_r \mu_{s^2}-(\nu_r \mu_s)^2\Delta t \Big) T_w = T_w \nu_r \Big( \mu_{S^2} - \nu_r\Delta t (\mu_S)^2 \Big),$$
which is the same as equation \eqref{eqn:varSever_pp}. 

This property will be subsequently used to determine variance in more complex cases, such as when we consider models with time-dependent probability of loss events.

\subsubsection{Summary of Homogeneous Poisson Process results}

Fig \ref{fig:homoCompare} demonstrates the accuracy of our analytic calculations for a wide variety of severity distirbution 
parameters.  The dark blue stars are the loss statistics in small time windows $\Delta t$ calculated by Monte Carlo, 
while the analytic formulas are solid black curves. The light blue stars are the loss statistics in a larger time window 
($T_w=1$ year here), following the dark/light blue coloring convention in Fig \ref{fig1:setup}.  Overall we see that 
our formulas are very accurate (see legend for further details).

The formulas we derived for the mean and variance of aggregate loss distributions assuming a 
homogeneous Poisson Process frequency distribution and independent severity distribution (Eq \eqref{eqn:meanR_homo}, \eqref{eqn:varR_homo}, 
\eqref{eqn:mnSever_pp}, \eqref{eqn:varSever_pp}) are straightforward calculations but have scarcely been reported.  
This could be due to a number of factors, including: focus on tail-end loss distribution, desire to keep methods 
proprietary, reliance on Monte Carlo simulations, lack of temporal dependence, etc. Nevertheless, our work shows 
analytic calculations are possible for this common frequency distribution model and very accurate. 

\section*{Declaration of Interest}

The authors report no conflicts of interest. The authors alone are responsible for the content and writing of the paper.


\begin{thebibliography}{}

\bibitem[\protect\citeauthoryear{??}{orx}{2020a}]{orx_url}
 (2020 (accessed August 25, 2020)a).
\newblock {\em ORX}.
\newblock \url{https://managingrisktogether.orx.org/about}.

\bibitem[\protect\citeauthoryear{??}{orx}{2020b}]{orx_urlData1}
 (2020 (accessed August 25, 2020)b).
\newblock {\em ORX}.
\newblock
  \url{https://managingrisktogether.orx.org/loss-data/annual-banking-loss-report}.

\bibitem[\protect\citeauthoryear{??}{orx}{2020c}]{orx_urlData2}
 (2020 (accessed August 25, 2020)c).
\newblock {\em ORX}.
\newblock
  \url{https://managingrisktogether.orx.org/orx-news/monthly-top-5-december-2019}.

\bibitem[\protect\citeauthoryear{Afonso, Mihov, Curti, et~al.}{Afonso
  et~al.}{2019}]{afonso19}
Afonso, G.~M., A.~Mihov, F.~Curti, et~al. (2019).
\newblock Coming to terms with operational risk.
\newblock ~(20190107), https://ideas.repec.org/p/fip/fednls/87301.html.

\bibitem[\protect\citeauthoryear{{Bank For International Settlements}}{{Bank
  For International Settlements}}{2014}]{bis_2006}
{Bank For International Settlements} (2014).
\newblock Basel committee on banking supervision. operational risk -- revisions
  to the simpler approaches.
\newblock ~{\em ISBN 978-92-9131-869-8}, https://www.bis.org/publ/bcbs291.pdf.

\bibitem[\protect\citeauthoryear{Barreiro and Ly}{Barreiro and
  Ly}{2017}]{BarreiroLy_RecrCorr_17}
Barreiro, A. and C.~Ly (2017).
\newblock When do correlations increase with firing rates in recurrent
  networks?
\newblock {\em PLoS Computational Biology\/}~{\em 13}, e1005506.
\newblock DOI: 10.1371/journal.pcbi.1005506.

\bibitem[\protect\citeauthoryear{Barreiro and Ly}{Barreiro and
  Ly}{2018}]{BarreiroLy_jmns_18}
Barreiro, A. and C.~Ly (2018).
\newblock Investigating the correlation-firing rate relationship in
  heterogeneous recurrent networks.
\newblock {\em Journal of Mathematical Neuroscience\/}~{\em 8}, 8.
\newblock DOI: 10.1186/s13408-018-0063-y.

\bibitem[\protect\citeauthoryear{Chernobai, Rachev, and Fabozzi}{Chernobai
  et~al.}{2008}]{chern08book}
Chernobai, A.~S., S.~T. Rachev, and F.~J. Fabozzi (2008).
\newblock {\em Operational risk: a guide to Basel II capital requirements,
  models, and analysis}, Volume 180.
\newblock John Wiley \& Sons.

\bibitem[\protect\citeauthoryear{De~Fontnouvelle, Jesus-Rueff, Jordan,
  Rosengren, et~al.}{De~Fontnouvelle et~al.}{2003}]{deFontn03}
De~Fontnouvelle, P., D.~Jesus-Rueff, J.~S. Jordan, E.~S. Rosengren, et~al.
  (2003).
\newblock Using loss data to quantify operational risk.
\newblock {\em Federal Reserve Bank of Boston\/},
  http://dx.doi.org/10.2139/ssrn.395083.
\newblock DOI: 10.2139/ssrn.395083.

\bibitem[\protect\citeauthoryear{Hawkes}{Hawkes}{1971}]{hawkes71}
Hawkes, A.~G. (1971).
\newblock Spectra of some self-exciting and mutually exciting point processes.
\newblock {\em Biometrika\/}~{\em 58\/}(1), 83--90.
\newblock DOI: 10.1093/biomet/58.1.83.

\bibitem[\protect\citeauthoryear{Hawkes}{Hawkes}{2018}]{hawkes18}
Hawkes, A.~G. (2018).
\newblock Hawkes processes and their applications to finance: a review.
\newblock {\em Quantitative Finance\/}~{\em 18\/}(2), 193--198.
\newblock DOI: 10.1080/14697688.2017.1403131.

\bibitem[\protect\citeauthoryear{Ji{\v{r}}ina}{Ji{\v{r}}ina}{2011}]{jivrina11}
Ji{\v{r}}ina, M. (2011).
\newblock Trends in banks operational risk losses.
\newblock In {\em Proceedings of the 10th WSEAS international conference on
  E-Activities}, pp.\  95--100. World Scientific and Engineering Academy and
  Society (WSEAS).
\newblock https://dl.acm.org/doi/abs/10.5555/2183339.2183354.

\bibitem[\protect\citeauthoryear{Kay}{Kay}{1993}]{kay1993}
Kay, S.~M. (1993).
\newblock {\em Fundamentals of statistical signal processing}.
\newblock Prentice Hall PTR.

\bibitem[\protect\citeauthoryear{Last and Penrose}{Last and
  Penrose}{2017}]{pplect17}
Last, G. and M.~Penrose (2017).
\newblock {\em Lectures on the Poisson process}, Volume~7.
\newblock Cambridge University Press.

\bibitem[\protect\citeauthoryear{Lindner, Doiron, and Longtin}{Lindner
  et~al.}{2005}]{lindner05}
Lindner, B., B.~Doiron, and A.~Longtin (2005).
\newblock Theory of oscillatory firing induced by spatially correlated noise
  and delayed inhibitory feedback.
\newblock {\em Physical Review E\/}~{\em 72\/}(6), 061919.
\newblock DOI: 10.1103/PhysRevE.72.061919.

\bibitem[\protect\citeauthoryear{Litwin-Kumar, Chacron, and
  Doiron}{Litwin-Kumar et~al.}{2012}]{litwin2012spatial}
Litwin-Kumar, A., M.~J. Chacron, and B.~Doiron (2012).
\newblock The spatial structure of stimuli shapes the timescale of correlations
  in population spiking activity.
\newblock {\em PLoS Comput Biol\/}~{\em 8\/}(9), e1002667.
\newblock DOI: 10.1371/journal.pcbi.1002667.

\bibitem[\protect\citeauthoryear{Litwin-Kumar, Oswald, Urban, and
  Doiron}{Litwin-Kumar et~al.}{2011}]{litwin2011balanced}
Litwin-Kumar, A., A.-M.~M. Oswald, N.~N. Urban, and B.~Doiron (2011).
\newblock Balanced synaptic input shapes the correlation between neural spike
  trains.
\newblock {\em PLoS Comput Biol\/}~{\em 7\/}(12), e1002305.
\newblock DOI: 10.1371/journal.pcbi.1002305.

\bibitem[\protect\citeauthoryear{McNeil, Frey, and Embrechts}{McNeil
  et~al.}{2015}]{mcneil15}
McNeil, A.~J., R.~Frey, and P.~Embrechts (2015).
\newblock {\em Quantitative risk management: concepts, techniques and
  tools-revised edition}.
\newblock Princeton university press.

\bibitem[\protect\citeauthoryear{Shea-Brown, Josi{\'c}, De~La~Rocha, and
  Doiron}{Shea-Brown et~al.}{2008}]{shea2008correlation}
Shea-Brown, E., K.~Josi{\'c}, J.~De~La~Rocha, and B.~Doiron (2008).
\newblock Correlation and synchrony transfer in integrate-and-fire neurons:
  basic properties and consequences for coding.
\newblock {\em Physical review letters\/}~{\em 100\/}(10), 108102.
\newblock DOI: 10.1103/PhysRevLett.100.108102.

\end{thebibliography}

\end{document}